\documentclass[fleqn]{aa}  
\pdfoutput=1
\usepackage{catchfile}
\newcommand{\getenv}[2][]{%
 \CatchFileEdef{\temp}{"|kpsewhich --var-value #2"}{\endlinechar=-1}%
 \if\relax\detokenize{#1}\relax\temp\else\let#1\temp\fi}
\getenv[\HOME]{HOME}%\show\HOME
\getenv[\PWD]{PWD}%\show\PWD
\usepackage{graphicx,color}
\usepackage[fleqn]{amsmath} 
\usepackage[varg]{txfonts}
\usepackage{natbib,xspace}
\usepackage{stmaryrd,marvosym}
\usepackage[switch]{lineno}

\newcommand{\del}{\ensuremath{\dfrac{\partial}{\partial t}}}
\renewcommand{\div}{\ensuremath{\vec{\nabla} \cdot}}

\newcommand{\rhb}[1]{\left\llbracket #1 \right\rrbracket }
\newcommand{\const}{\ensuremath{\mathrm{const.}}}
\renewcommand{\dot}[1]{\stackrel{.}{#1}}
\renewcommand{\hat}[1]{\stackrel{\wedge}{#1}}
\renewcommand{\tilde}[1]{\ensuremath{\stackrel{\sim}{#1}}}

\newcommand{\Ms}{M_{1}^{2} \sin^{2}\vartheta}

\renewcommand{\deg}{\ensuremath{^{\circ}}}
\newcommand{\SA}{shock angle\xspace}
\newcommand{\SAs}{shock angles\xspace}
\newcommand{\FA}{flow deflection angle\xspace}
\newcommand{\FAs}{flow deflection angles\xspace}
\newcommand{\LamC}{$\lambda$ Cephei\xspace}
\renewcommand{\H}{\ensuremath{\mathrm{H}}}
\newcommand{\He}{\ensuremath{\mathrm{He}}}
\newcommand{\p}{\ensuremath{\mathrm{p}}}
\newcommand{\PUI}{\ensuremath{\mathrm{PUI}}}
\newcommand{\Ostar}{O star\xspace}
\newcommand{\Ostars}{O stars\xspace}
\newcommand{\sfl}{single-fluid\xspace}
\newcommand{\lsc}{large-scale\xspace}
\newcommand{\Lsc}{Large-scale\xspace}
\newcommand{\ide}{in-depth\xspace}
\newcommand{\Ha}{H$\alpha$\xspace}

\begin{document}

\title{ Shock structures of astrospheres}
\author{K. Scherer \inst{1,2} \and H. Fichtner \inst{1,2} \and
J. Kleimann \inst{1} \and T. Wiengarten \inst{1}\and D.J. Bomans \inst{2,3} \and K. Weis \inst{3}}
\institute{Institut f\"ur Theoretische Physik IV: Weltraum-
  und Astrophysik, Ruhr-Universit\"at Bochum, 44780 Bochum, Germany,
\email{kls@tp4.rub.de, hf@tp4.rub.de}
\and
 Research Department, Plasmas with Complex Interactions, Ruhr-Universit\"at Bochum, 44780 Bochum, Germany 
\and
Astronomisches Institut, Ruhr-Universit\"at Bochum, 44780 Bochum, Germany, \\
 \email{bomans@astro.rub.de}
}

\date{Received: accepted}

\abstract
{The interaction between a supersonic stellar wind and a (super-)sonic
  interstellar wind has recently been viewed with new interest. We
  here first give an overview of the modeling, which includes the
  heliosphere as an example of a special astrosphere. Then we
  concentrate on the shock structures of fluid models, especially of
  hydrodynamic (HD) models. More involved models taking into account
  radiation transfer and magnetic fields are briefly sketched. Even
  the relatively simple HD models show a rich shock structure, which
  might be observable in some
  objects. %Coupling the neutral interstellar gas flow into the models
%  will lead to remarkably different results.  
}
{We employ a \sfl model to study these complex shock
  structures, and compare the results obtained including heating and cooling
   with results obtained without these effects. Furthermore, we
  show that in the hypersonic case valuable information of the shock
  structure can be obtained from the Rankine-Hugoniot equations.  }
{We solved the Euler equations for the \sfl case and also for
  a case including cooling and heating. We also discuss the
  analytical Rankine-Hugoniot relations and their relevance to
  observations.  }
{We show that the only obtainable length scale is the termination shock
  distance. Moreover, the so-called thin shell approximation
  is usually not valid. We present the shock structure in the model
that  includes heating and cooling, which differs remarkably from that of
  a \sfl scenario in the region of the shocked interstellar
  medium. We find that the heating and cooling is mainly important in
  this region and is negligible in the regions dominated by the stellar
  wind beyond an inner boundary.  }
{}

   \maketitle

\section{Introduction}

Simulations of astrospheres around hot stars have recently moved into
the focus of scientific research, see, for example,
\citet{Decin-etal-2012}, \citet{Cox-etal-2012}, \citet{Arthur-2012},
and \citet{van-Marle-etal-2014}. These authors modeled astrospheres using
a (magneneto-)hydrodynamic approach, either in one or two dimensions (1D
or 2D). Especially \LamC is an interesting example; this is the brightest runaway \Ostar in the sky (type O6If(n)p) and is used
as an example for studying the shock structure in \sfl hydrodynamical
models with and without cooling and heating.  Runaway O and B stars
are common and part of a sizable population in the Galaxy. A
significant number of these show a bow-shock structure
\citep[e.g.,][]{Huthoff-Kaper-2002,Gvaramadze-Bomans-2008,Gvaramadze-etal-2011,Kobulnicky-etal-2010}.
The survey conducted by \citet{Peri-etal-2012} and
\citet{Peri-etal-2015} recently listed many bow shocks around \Ostars and 
the corresponding Herschel observations in the infrared
\citep{Cox-etal-2012}.  Other OB stars also show supersonic
relative velocities with respect to the ambient interstellar
medium (ISM)
\citep{Povich-etal-2008,Sexton-etal-2015}. From a modeling point of
view, it is irrelevant whether the star or the ISM is in supersonic
motion, it is only of interest that the relative speed between the two is
supersonic. The latter condition is required for the formation of a
bow shock in hydrodynamics. When a magnet field is included, the inflow speed
has to be higher than fast magnetosonic to create a bow shock.

Hot stars are not alone in developing shock structures, cool F, G, and even
M stars are driving supersonic winds, and with a supersonic relative
speed with respect to the ambient ISM, show bow shocks. Some of these
structures of nearby stars can be observed in Lyman-$\alpha$
lines. The latter are excited when neutral hydrogen atoms enter the
shock region and are slowed down (see below)
\citep{Wood-etal-2007,Linsky-Wood-2014}.

We here model the interaction of all these stars with the surrounding ISM
by the Euler equation with various extensions (see
Sects.~\ref{fluideq} and~\ref{specialfluid}). We assume throughout that the stellar wind is supersonic far away from the
star. The relative speed between the ISM and the star can be
supersonic or subsonic. In the latter case, no bow shock is generated,
therefore a supersonic
relative
motion is required for modeling or observing bow shocks.

We assumed the stellar wind plasma in all simulations to be
``collisonless'' and hence the shocks are also collisionless
(``collisionless shocks''). Therefore, the plasma and the neutral
fluid are coupled only through charge exchange, electron impact, or
photoionization processes \citep[e.g.,\ ][]{Scherer-etal-2014}, which lead to a
mass-, momentum-, and energy loading of the plasma fluid because the
respective bulk velocities in the shocked regions are different from
those inside the termination shock.

In the following, we briefly discuss the Euler equations needed for modeling the interaction between stellar winds and the ISM,
including different extensions (such as magnetohydrodynamics, cooling and
heating, and multifluid approaches). Then we concentrate on
modeling $\lambda$ Cephei and discuss some possible future
applications for observations.

In Sect.~\ref{fluideq} we give a general overview of applicable
conservative fluid models, which is followed in
Sect.~\ref{specialfluid} by a short discussion of the different
fluid-based modeling approaches, which will be subject of forthcoming
publications. In Sect.~\ref{single} we discuss the general shock
structure in a \sfl hydrodynamic model without other
interactions. There we present some crucial features, which to our
knowledge have not appeared in earlier publications. In
Sect.~\ref{cool} we compare the model discussed in
Sect.~\ref{single} with a model that includes cooling and heating functions (CHFs),
which is then followed by a summary.

\section{Fluid equations\label{fluideq}}

In a series of consecutive papers we intend to discuss different models
for runaway (O) stars, for which \LamC (type O6If(n)p) is used as an
example \citep{Huthoff-Kaper-2002}. In this first
paper we discuss a \sfl hydrodynamic model (pure hydrodynamic
model: pH model) and another model that includes cooling and heating
functions (CH model). In \citet{Scherer-etal-2015a} we already used the
CH model to estimate the cosmic ray flux in \LamC. Here we discuss the underlying shock structure and
some observational consequences derived from the Rankine-Hugoniot
relations in greater detail.

The plasma in and around astrospheres is usually highly dilute, so that
the standard fluid description seems to fail for the characteristic
length scales $L$ that are commonly used (between several hundreds of AU for cool stars and
(tens of) parsecs for hot stars). The reason is that the mean free
path (mfp) $\lambda_{coll}= 1 / (\sigma_{T} n)$ for hard sphere
collisions that is typically assumed in fluid theory is on the same order
as $L$: Using the Thompson cross section $\sigma_{T} =
10^{-15}$\,cm$^{2}$ and a plasma density of $n=10^{-3}$\,cm$^{-3}$, which is usually that
in front of the termination shock, we obtain
$\lambda_{coll}=10^{18}$\,cm~$\approx 1$\,pc.  For ISM densities
in the range between $0.1$ to 100\,cm$^{-3}$ the mfp is on the order
of thousands of AU to 1\,AU. Nevertheless, this collision mfp is only
valid for neutral fluids because the collision between
Debye spheres has to be taken into account in plasmas. The Debye radius
$\lambda_{D}$ of a Debye sphere is
$\lambda_{D} \approx 6.9 \sqrt{T[k]/n[cm^{-3}]}$\,cm. For plasma
temperatures between a few hundred to a billion K after the
termination shock and the above
discussed densities, we find $\lambda_{D}$ in the range of a few meters
to kilometers. This means that the plasma fluid is dominated by the collisions
between Debye spheres instead of hard spheres. The plasma is therefore
collisionless and the resulting shocks are collisionless
shocks. Only in this case can ionized fluids be described as collisionless,
the neutral fluids have mfps that are much longer. The neutrals are
sometimes handled kinetically, but see \citet{Heerikhuisen-etal-2006}
and \citet{Alouani-Bibi-etal-2011} for a comparison between fluid and
kinetic models.

Nevertheless, the interaction between the neutral fluid and the plasma
through charge exchange processes leads to some important modifications of
the fluids: As a result of the mass- and momentum loading, the plasma fluid
slows down, and  the original plasma is
heated through the energy loading. These processes are well studied in the heliosphere
\citep[][]{Fahr-etal-2000,
  Pogorelov-etal-2006,Alouani-Bibi-etal-2011,Scherer-etal-2014}.

Another complication arises from the fact that hot stars completely
ionize the ambient ISM up to a radius $r_{S}$, the Str\"omgren
radius. In the earlier days of heliospheric science, this lead to some
confusion because $r_{S}$ for the heliosphere is on the order of
1000\,AU, but interstellar neutrals were observed at 1\,AU
\citep[for a review see][]{Fahr-1990}. There are also recent attempts
\citep{Mackey-etal-2013a,Mackey-etal-2014b} to model the ionization
fronts around hot stars with a relative motion to the ambient ISM and
for Str\"omgren spheres \citep{Ritzerveld-2005}. All these arguments
lead to the conclusion that neutrals are an important feature in
modeling astrospheres, even around hot stars.

Moreover, \LamC is observed in \Ha light
\citep{Scherer-etal-2015a}, which means that many neutrals are excited inside the bow shock region. This holds true
for all stars with bow shocks observed in \Ha. Because of the
increasing photon flux toward the star, the neutrals can become
totally ionized in the astrosheath and may not reach the inner
astrosphere, so that the unshocked stellar wind is not
affected. In the outer astrosheath, were the \Ha is
observed, however, the neutrals have to be self-consistently taken into
account. 

As stated above, from a modeling point of view, only the relative
velocity between the star and the ambient ISM is of interest. We therefore here transform all data into a stellar-centric coordinate
system throughout, such that the star is at rest, the x-axis points in the
interstellar plasma flow direction, the y-axis is perpendicular to it
in the plane of inflow, and the z-axis is orientated in such a way as to
complete the right-handed system. Thus the inflow is always parallel
to the x-axis.

The inflow direction is also called ``upwind'', while the opposite
direction is called ``downwind''. This has to be clearly distinguished
from the ``upstream'' direction which, in the shock restframe, denotes
the incoming supersonic direction, while ``downstream'' is the
outflowing subsonic direction.  

To set the theoretical basis for all the models, we use
Eq.~(\ref{eq:1}) below in various simplifications. First we discuss a very
general set of fluid equations. Then we concentrate in this work on
the \sfl approach with and without cooling and heating, but
will use the general set for future work.

A variety of (multi-) fluid (Euler) equations with
different extensions is used in the literature
\citep[see e.g.\
][]{Pogorelov-etal-2009,Bouquet-etal-2000,
  Fahr-etal-2000,Jun-etal-1994,Downes-Drury-2014}.
The set of Euler equations (the continuity, momentum, and energy
equation) can be combined into
\begin{eqnarray}\nonumber
  \label{eq:1}
  \del \begin{bmatrix} \rho_{j} \\ \rho_{j} \vec{v}_{j} + \mathcal{P}_{1}
    \vec{F}_{rad} \\ E_{j} + \mathcal{P}_{2} E_{rad}\\\vec{B} \end{bmatrix} 
+\div   \begin{bmatrix}
        \rho_{j} \vec{v}_{j} \\
        \rho_{j} \vec{v}_{j}\vec{v}_{j} + P_{j} \widehat{I} +
        \mathcal{P}_{3} \vec{F}_{rad} - \dfrac{\vec{B}\vec{B}}{4\pi}\\[0.25cm]
        (E_{j} + P_{j} )\vec{v}_{j} + \mathcal{P}_{4} \vec{F}_{rad}
        -\dfrac{\vec{B}(\vec{B}\cdot\vec{v_{j}})}{4\pi} \\
        \vec{v}_{j}\vec{B}-\vec{B}\vec{v}_{j}
         \end{bmatrix} =\\
  \begin{bmatrix} 0 \\ \rho_{j} \vec{F} + \div \widehat{\sigma} -
    \vec{\nabla} P_{CR}
   \\
      \rho_{j} \vec{v}_{j}\cdot\vec{F} + \div (\vec{v}_{j}\cdot\widehat{\sigma}) -
      \div \vec{Q} - R_{L} -\vec{v}_{j}\cdot\vec{\nabla} P_{CR}\\0\end{bmatrix}
+ \begin{bmatrix} S_{j}^{c} \\ \vec{S}_{j}^{m}\\S_{j}^{e}\\\vec{A}\end{bmatrix}
,\end{eqnarray}
where the left-hand side is written in a conservative form.
$\vec{v}_{j}, \rho_{j}, E_{j}, P_{j}$ are the fluid velocity, mass density,
total energy density, and pressure of species $j$, $\widehat{I}$ is the
unit tensor, $\widehat{\sigma}$ the viscosity or stress tensor, $\vec{F}$
is an external force per unit mass and volume, $\vec{Q}$ is the heat
flow, $S_{j}^{r}$ are sources and sinks caused by charging processes
in the continuity and energy equation $r\in\{c,e\},$ and
$\vec{S}_{j}^{m}$ those of the momentum equation.  $R_{L}$ is a
cooling function. The parameters with the subscript $_{rad}$ describe
the momentum ($\vec{F}_{rad}$) and energy ($E_{rad}$) coupling to the
radiation transport, and $P_{CR}$ the coupling to cosmic rays. The
$\mathcal{P}_{k}, k\in\{1,2,3,4\}$ are constants. Finally,
$\vec{A}$ describes the ambipolar diffusion between the neutrals
and the ions.

\begin{table}[t!]
  \centering
  \begin{tabular}{l|l}
    $\dot{M}$ & $1.5\cdot10^{-6}$ $M_{\odot}$/yr\\
    Terminal speed  & 2500 km/s\\
    Spectral type & O6If(n)p \\
    Distance  & 649$^{+112}_{-63}$ pc\\
  \end{tabular}
  \caption{Parameters for \LamC
    \citep{van-Leeuuwen-2007}.}
  \label{tab:1}
\end{table}

The equations for species $j$ (ion species $j_{i}$ and neutral species
$j_{n}$) are composed of the governing equations
\citep{Scherer-etal-2014} for the sum of all ions ($\sum j_{i}=I$) and
neutrals ($\sum j_{n}=N$), which are coupled through the source terms
$S^{c,e}$ and $\vec{S}^{c}$ and balance equations for single species,
like $i\in\{\p, \He^{+}, \He^{++},\PUI ...\}$ for the ions and
$n\in\{\H, \He, ...\}$ for the neutrals, and in addition by ambipolar
diffusion $\vec{A}$. Most of the contemporary heliospheric models
include neutral hydrogen and pickup ions, which behave differently
from the original stellar wind ions
\citep{Fahr-etal-2000,Scherer-Ferreira-2005a,
  Pogorelov-etal-2009,Alouani-Bibi-etal-2011}.
For models including helium, see \citet{Izmodenov-etal-2003a}.
The governing equations describe the dynamic behavior, while the balance
equations determine the density and energy density (temperature) of
the considered species.

In the following we use the single-fluid module of the CRONOS code
\citep{Kissmann-etal-2008,Kleimann-etal-2009,
  Wiengarten-etal-2015,Scherer-etal-2015a}.
To enable direct comparisons with future models that include the magentic
field, multifluids, and other complications, we performed the calculations
in a 3D spherical coordinate system.

\subsection{Ion-neutral interactions}

As stated above, the bow shock region of \LamC is observed in \Ha
light. The latter can be generated either by neutral H-atoms
penetrating the astrosphere, or by recombination. In both cases
the neutral gas is not directly affected by the plasma flow, only
indirectly by charge exchange processes. The latter can lead to severe
changes in the plasma fluid dynamics because the interstellar neutral
fluid and the plasma usually have different bulk speeds. The situation
is more complicated when the neutral gas is produced by recombination,
because then it locally has the same bulk speed as the plasma, but this is not true
globally because the neutrals are not affected by
electromagnetic interactions. These charge exchange processes are well
studied for the heliosphere, and because they are also applicable to
astrospheres, we briefly describe them below.  For recent heliospheric
observations and their interpretation see, for example,\
\citet{Bzowski-etal-2015}, \citet{McComas-etal-2015b}, and
\citet{Sokol-etal-2015}.

The total ion and neutral densities are the sum of the respective
partial densities $\rho_{I}=\sum_{j_{i}}\rho_{j_{i}}$ and
$\rho_{N}=\sum_{j_{n}}\rho_{j_{n}}$.  The total thermal energies
$E_{I}, E_{N}$ and pressures $P_{I}, P_{N}$ are the sum of the partial
thermal energy densities
$E_{I}= \sum_{j_{i}} E_{j_{i}}, E_{N}= \sum_{j_{n}} E_{j_{n}}$ and
partial pressures
$P_{I}= \sum_{j_{i}} P_{j_{i}},P_{N}= \sum_{j_{n}} P_{j_{n}}$. In
the most general case the velocities $\vec{v}_{I}, \vec{v}_{N}$ in the
governing equations can be functions of the velocities in the balance
equations. Here we assume that all processes that induce different
partial velocities $\vec{v}_{j}$ reach equilibrium velocities
$\vec{v}_{I}$ and $\vec{v}_{N}$ in the governing equations on timescales much shorter than the fluid timescales. In this sense, the
velocities in the balance equations are identical to those in the
governing equations for the ions and neutrals, respectively. We also
assume that the induction equation only depends on $\vec{v}_{I}$.

A balance equation of a given species $j$ describes the variation of
its mass 
density $\rho_{j}$ and energy density $E_{j}$ (or pressure
$P_{j}$). The charging processes, such as charge exchange between a
neutral and an ionized particle, electron impact, and photo-ionization,
change the density $\rho_{j}$  and pressure $P_{j}$ (or $E_{j}$) of
that species and through the sum of the governing equations. The momentum
change is described in the governing equations. A newly charged
``neutral'' of species $n_{s}$ will, after being immediately picked up
by the stellar magnetic field, be removed from the density
$\rho_{n_{s}}$ and become an ion contributing to the ion density
$\rho_{i_{s}}$. In case of charge exchange between a neutral species $n_{s}$
and an ion species $i_{r}$, an additional energetic neutral of species $r$ can
be produced \citep[for details see][]{Scherer-etal-2014}, which should
be treated as a separate species because it also has a different energy
density than the original neutral of species $s$. Thus the
balance equation for each species describes the temporal and spatial
behavior of its partial density and partial pressure, which then are
added to give the total density and pressure.

The momentum change is only calculated in the governing equation
because all neutrals and ions flow, by assumption, with the respective
bulk velocity $\vec{v}_{N},\vec{v}_{I}$. Therefore no momentum equation for a single
ion or neutral species is needed.

The charging processes lead to a mass, momentum, and energy loading
of the neutral and ion fluids that change the overall dynamics.
For an overview of the relevance of charge exchange, electron
  impact and photoionisation processes on the fluid dynamics see
  \citet{Scherer-etal-2014}.

  These processes are independent of the
  relative velocity between the star and the ISM only at first glance. The
  reason for the dependence is that the neutrals can deeply penetrate the
  astrosphere when the relative motion between the star and the ISM
  is high enough, as can be seen in the example of the heliosphere
  (see discussion above). Thus a self-consistent model of an
  astrosphere needs to include the neutral fluids as well as those of
  the newly created ions because of their characteristics, such
as   velocity and mass.

Another complication is that neutrals can also be
  produced by recombination inside the bow shock. The timescale $t_{rec}$ for recombination is
  \begin{eqnarray}
    t_{rec} = \frac{1}{\beta_{2} n} \approx 10^{11}s
  ,\end{eqnarray}
where $\beta_{2}\approx 2\cdot 10^{-10}$\,cm$^{3}$/s is the rate
coefficient needed to recombine to the second energy level, and
$n=44$\,cm$^{-3}$ is the density between the bow shock and the
astropause (see below), which is higher than the timescale $t_{ph}$
for photoionization
\begin{eqnarray}
  t_{ph} = \frac{1}{v n \sigma_{ph}} \approx 6\cdot 10^{9} s
,\end{eqnarray}
with the velocity $v=40$\,km/s and the photoionization cross-section
$\sigma_{ph} \approx 10^{-17}$\,cm$^{2}$  . Therefore, the recombination
should not affect the ionization state of the plasma. Nevertheless,
an energy loss
or gain
occurs during the recombination and ionization process  because of the involved photons. This process is included through the
cooling and heating functions.

While the latter process only affects the energy equation, the
penetrating neutrals from the ISM affect the continuity equation (for
different masses such as hydrogen and helium) and the momentum equation
as well as the energy equation.

\subsection{Heating and cooling}

Conductive heat transfer is taken into account mainly in models
concerning the undisturbed solar wind from around 100 solar radii to
the termination shock (TS) \citep{Usmanov-Goldstein-2006}. The problem
with these models is that the electrons are important for the heat transfer. The behavior of the electrons is not well understood at the
termination shock and beyond, but see
\citet{Chashei-Fahr-2013,Chashei-Fahr-2014}.  For \Ostar astrospheres,
\citet{Arthur-2007} included a heat transfer model to avoid too strong
adiabatic cooling of the stellar wind at the TS.

Including radiative cooling or heating effects is especially important
in the outer astrosheath region, that is,\ between the bow shock (BS) and
the astropause (AP), because here the shocked ISM is
dense and hot, in contrast to the shocked stellar wind, which is
dilute, but extremely hot in the region between the TS and the AP. The
cooling is caused by inelastic collisions, in which an atom is
excited. After the electron returns to a lower energy state, it
re-emits a photon that carries the energy away. This cooling
then leads to a collapse of the outer astrosheath. For more details see
below.

The high photon flux of hot stars means that a radiative heating function
\citep[like that in][]{Kosinski-Hanasz-2006} for the stellar wind and
the ambient ISM needs to be included because it can
change the state of the ambient ISM.

Details of cooling and heating functions (CHFs) are discussed in
Sect.~\ref{cool}.

\subsection{Rankine-Hugoniot relations}

Form the conservative form of the Euler equations it is easy to obtain
the Rankine-Hugoniot relations by substituting
$\partial_{t} f \rightarrow -u \rhb{f}$ and
$\nabla f \rightarrow \vec{n} \rhb{f}$
\citep[e.g.,\ ][]{Goedbloed-Poedts-2004}, where $f$ is the quantity
that jumps at the shock, $u$ is the shock speed, and $\vec{n}$ the
shock normal. The bracket $\rhb{f}$ is defined as
$\rhb{f} \equiv f_{1}-f_{2}$, where the index $1$ denotes the state
upstream of the shock, while $2$ describes the state downstream of the shock. The
Rankine-Hugoniot relations can be simplified by transforming them into
the shock rest frame, or by assuming that the shock has become stationary,
which we assume to hold in the following
\citep[see, e.g.,\ ][]{Naca-1135}. We note that the dynamical shock
interactions can be different
\citep[see ][]{Courant-Friedrichs-1948,Edney-1968,Emanuel-2000}.

In Sect.~\ref{specialfluid} we give a short overview of shocks for special
conditions for Eq.~(\ref{eq:1}). This is followed by an \ide analysis of
astrospheric hydrodynamic \sfl shocks
(Sect.~\ref{single}). In Sect.~\ref{cool} we discuss the
influence of heating and cooling.

\subsection{Polytropic index}

The ideal gas equations are commonly assumed together with a polytropic
equation of state ($P/\rho^{n}=$const) as closure for the above
equation system~\ref{eq:1}. For a single-fluid of a mono-atomic gas
without heat transfer, the polytropic index $\gamma=5/3$
is often used instead of the polytropic index $n$. This is also our choice
throughout the manuscript. 

The ISM can still contain molecules with different $\gamma$s,
and heat transfer may also play a role after passage through a
shock. Moreover, molecules can dissociate during or after the shock
passage, which changes the polytropic index. In a single-fluid model
different polytropic indices also lead to different compression ratios
of the shocks and to varied extents of the regions between the
astro- or heliopause and the bow or termination shock.

A thorough discussion of this topic is given in
\citet{Scherer-etal-2015c}, but see also \citet{Izmodenov-etal-2014}.

\section{Special fluids\label{specialfluid}}

Setting in Eq.~(\ref{eq:1}) 
$\mathcal{P}_{i}=0, \vec{B}=\vec{0}, P_{CR}=0, \widehat{\sigma}_{ij}=0,
\vec{F}=\vec{0}, \vec{A} = \vec{0}, S^{m,e}=0, \vec{S}^{m}=\vec{0}$,
and $R_{L}=0,$  we obtain the pure hydrodynamic case. 

The discussion below holds true for mono-atomic gases.  The
high temperatures after the shock passage can cause di-atomic or molecular gases
to dissociate, which can change the flow dynamics remarkably. These
gases are not discussed here.

\subsection{Radiative hydrodynamic shocks}
\label{sec:rdh}
Allowing $\mathcal{P}_{i}> 0$, the radiative hydrodynamic (RHD)
scenario is set up, for example, in \citet{Lowrie-etal-1999}. The RHD
shocks are discussed in \citet{Bouquet-etal-2000}, where the authors
noted that the compression ratio $s$  in the non-relativistic case can also reach values of up to $s=7$, in contrast to pure hydrodynamic shocks,
where the compression ratio is  $s=4$ at most. In the literature there
is a distinction  between ``radiative shocks'' and RHD, the former are
characterized by the fact that only the post-shocked plasma (gas) is
efficient cooled by radiation, while in RHD the first and second
moments of the radiation transfer equation are included. The latter
case is more general because it includes the former if the radiation
transfer moments are negligible in the pre-shocked plasma (gas).

The radiative shock scenario is commonly described by including
cooling and heating functions in the energy equation. The RHD
treatment applies when the optical depth is great, that is, when\ the photon
mean free path is shorter then the characteristic dimensions $L$.

\subsection{Shocks modified by cosmic rays}
The influence of the cosmic rays ($P_{CR}\ne0$) has been discussed for
the heliosphere by \citet{Fahr-etal-2000},
\citet{Alexashov-etal-2004}, \citet{Zank-Frisch-1999},
and \citet{Florinski-etal-2004}, among others. For cosmic-ray-driven shocks
on Galactic scales we refer to \citet{Salem-Bryan-2014} and references
therein, and to \citet{Blasi-2013} for a general overview. The
compression ratio can also become higher than four.

\subsection{Magnetohydrodynamical shocks} 

Including only the magnetic field $\vec{B}\ne \vec{0}$ and neglecting all
other effects
($\mathcal{P}_{i}=0, p_{CR}=0, \widehat{\sigma}_{ij}=0, \vec{F}=\vec{0},
\vec{A}_{j} = \vec{0}, S^{m,e}=0, \vec{S}^{m}=\vec{0}$,
$R_{L}=0$) one is lead to \sfl magnetohydrodynamics (MHD).  MHD
shocks are much more complicated than pure HD shocks
because of the additional characteristic wave speeds: the normal
Alfv\'en speed, and the normal fast and slow magnetosonic speeds. This
allows for different shock types, for example, slow and fast shocks,
where the magnetic field is increased (fast shock) or diminished (slow
shocks). Other types of shocks are possible, for instance, intermediate shocks and
switch-on or switch-off shocks. For an \ide view see
\citet{Goedbloed-2008} and \citet{Goedbloed-etal-2010}, and for an
application to the heliosphere see
\citet{Scherer-Fichtner-2014}. \Lsc MHD models for the
heliosphere can be found, for example, in \citet{Pogorelov-etal-2013}
and \citet{Opher-etal-2012} and references therein.

\subsection{Fluids including the stress tensor or forces}

Internal and external forces $\vec{F}$, such as buoyancy, drag, lift,
tides, and gravitation, are often neglected in collisionless
astrospherical plasmas. Although  these forces can become important
\citep{Usmanov-etal-2014} at the boundary between the fast
and
slow stellar wind, we set all forces
$\vec{F}=\vec{0}$ and the stress tensor $\widehat{\sigma}_{ij}=0$ to
zero for all practical purposes.

To our knowledge, the cases
$\widehat{\sigma}_{ij}\ne0, \vec{F}\ne \vec{0}, \vec{A}_{j}\ne
\vec{0}$
have not yet been discussed in the context of
astrospheres.  In principle, the stress
tensor $\widehat{\sigma}_{ij}=0$ is needed to physically induce
instabilities, which is typically simulated by ``numerical'' viscosity,
because otherwise the required resolution is too high to be modeled.

\subsection{Fluids including dust}

Many bow shocks are observed in infrared radiation emitted by dust
grains \citep{Cox-etal-2012}. \citet{van-Marle-etal-2011} modeled
the dynamics of the dust grains generated by the star. Far away from
the star, neither the gravitational force nor the radiation pressure
plays a role because both decay as $r^{-2}$. Furthermore, these
authors treated the dust as test particles: The dust
particles experience a drag force by the plasma, but the plasma is
not affected by the dust. This behavior is also found in the solar
wind \citep{Fahr-etal-1995}. 

Nevertheless, the dust dynamics are complicated and can be modeled
using the Euler equation when external forces such as drag,
Poynting-Robertson forces, and the Lorentz force are included. 

In principle, dust grains can affect the plasma by sputtering of
ions or atoms from their surface, which leads to mass, momentum, and energy
loading of the plasma. To model this, additional balance equations are
needed, as is a dust model, to describe their dynamics, which are
important for the momentum and energy loading. With a gas-to-dust mass
ratio of about 100 \citep{Liseau-etal-2015}, the plasma dynamics should
not be affected by the dust, and we therefore did not take it into account
here. From an observational point of view, especially using
infrared observations from Herschel \citep{Cox-etal-2012}, the dust is
important for detecting bow shocks.

\section{Hydrodynamic shocks: \sfl\ shocks\label{single}}

We here assume that the relative motion between the star
and the ISM is supersonic (for subsonic relative
speeds see \citet{Steinolfson-1994} and \citet{Zank-etal-2013}), and
furthermore, that the shock structures are in quasi-stationary
equilibrium, which means that they are at rest with respect to the star.

Now we concentrate on \sfl hydrodynamic shocks, where we
assume that the fluid is collisionless, consists of protons, and is
polytropic with a coefficient $\gamma = 5/3$. Such a fluid behaves like
a neutral perfect gas fluid, and all the gas dynamical methods can be
applied \citep{Ben-Dor-2007}. For \lsc simulations of
astrospheres see \citet[][]{Arthur-2012}, and for
heliosphere simulations see \citet{Pauls-etal-1995}, \citet{Fahr-etal-2000}, and
\citet{Izmodenov-etal-2003a}.

As pointed out in \citet{Baranov-etal-1971} and
\citet{Pauls-etal-1995}, the shock structure of the heliosphere
consists of strong shock solutions, where the shocked Mach number
$M_{2}<1,$ and weak solutions, where the Mach number after the shock is
higher than 1 ($M_{2}>1$), see also
\citet{Izmodenov-Baranov-2006}. Making the HD equation dimensionless
by setting
\begin{eqnarray}
  \label{eq:dim}
 \rho = \rho_{0}\tilde{\rho} \quad v = v_{0}\tilde{v} \quad t =
 T_{0}\tilde{t} \quad r = R_{0}\tilde{r} \quad P = \rho_{0}v_{0}^{2}\tilde{P}
,\end{eqnarray}
a factor $f=v_{s}t_{TS}/R_{TS}$ in front of the $\div()$ operator
appears, which includes the characteristic scales for the speed, time,
and length scales.  The factor is one by construction when the speed is set equal to the constant supersonic
stellar wind speed at the inner boundary $v_{0}=v_{s}$, the timescale $t_{0}$
 to the scale the stellar wind needs to reach the TS $T_{0}=t_{TS}$, and
length scale to the stellar-centric distance to the TS $R_{0}= R_{TS}$
on the stagnation line. The
dynamically independent values here are the stellar wind density
and speed as well as the stagnation distances, which are determined by
the total momentum conservation between the stellar wind and the
ISM (see below). Thus, all
 different astrospheres can be modeled after solving for the
dimensionless fluid and then scaling it back by using
$\rho_{0}, v_{S}$, and $R_{TS}$. 

The initial mass density $\rho_{0}$ and the terminal speed $v_{s}$ are
prescribed at the inner boundary. The stellar-centric TS distance
$R_{TS}$ is a function of the total momentum of the stellar wind and
the ISM (see below).

Therefore, we no longer distinguish between different astrospheres
\citep{Arthur-2007,Arthur-2012}, pulsar wind nebulae
\citep{Bucciantini-2002,Bucciantini-2014} or the heliosphere, as long
as we can describe them by a single collisionless fluid without other
interactions. In that sense, this section is generally applicable to all
\sfl astrospheres. Therefore, we use the dimensionless
parameters below, and to save writings we drop the $\tilde{}$
in what follows.

\subsection{\Lsc shock structure}
\begin{figure*}[t!]
  \centering
  \includegraphics[width=0.9\textwidth]{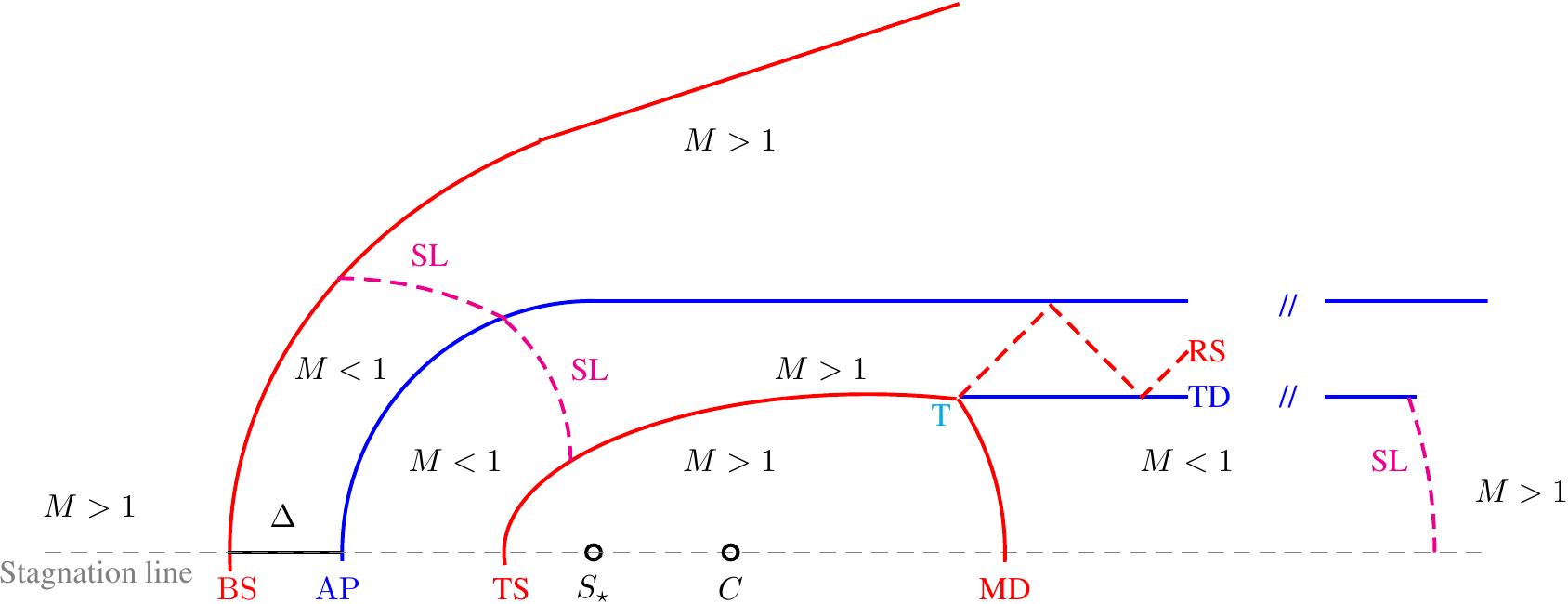}
  \caption{Sketch of a hydrodynamic astrospherical shock
    configuration. The red lines denote the shocks, the blue
    lines stand for tangential discontinuities. The Mach numbers in the
    different regions are indicated by $M\gtrless 1$.  BS denotes the
    bow shock, TS the termination shock, SL is the sonic line,
    $S_{\star}$ the position of the star, and $C$ the middle point of an
    radius $R_{M}$ describing the Mach disk MD. TD is the tangential
    contact discontinuity starting at the triple point $T$, where
    the reflected shock RS also starts.\label{rh:1}}
\end{figure*}

\begin{figure}[t!]
  \centering
  \includegraphics[width=0.7\columnwidth,angle=90]{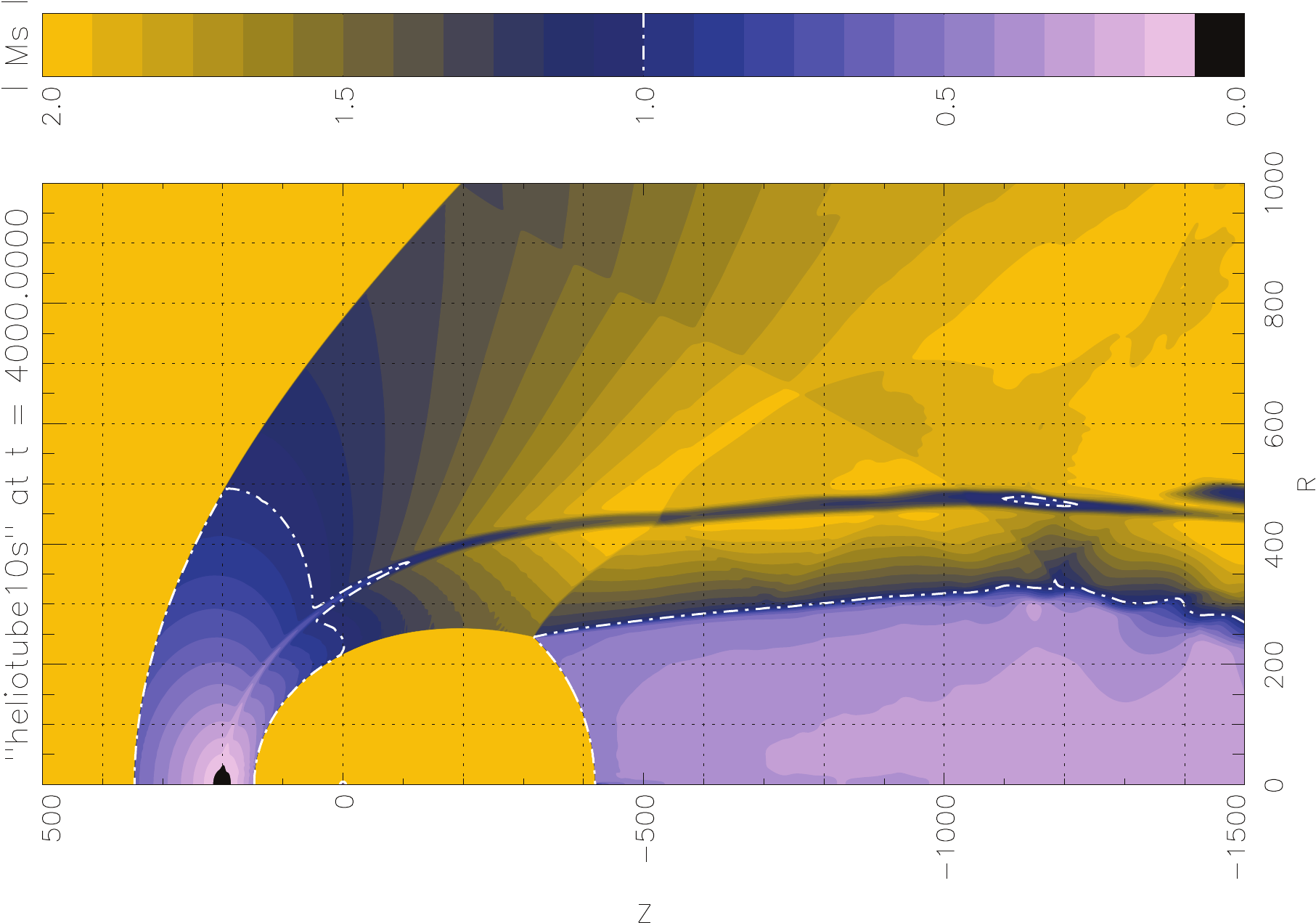}
  \caption{ Contour plot of the sonic Mach number $M$ from an polytropic
    one-fluid simulation. The features described in the text and shown
    in Fig.~\ref{rh:1} can easily be recognized. The kink at
    $x=-250$, and $R = 1000$ is caused by a reflection at the boundary of
    the integration region. Values are truncated at $M=2$.\label{rh:2}}
\end{figure}

The shock structure for a single collisionless fluid is sketched in
Fig.~\ref{rh:1}, and a model is shown in Fig.~\ref{rh:2}. 
 Most of the shock structures in a \sfl description, such as the TS
distance and the jump conditions, can be described analytically, or
semi-analytically (for example, the AP and BS distance). 
Therefore, they can be used for a first estimate, and we discuss them in
detail below.
 
In the hydrodynamic case each surface of flow lines can be replaced by a solid wall
without changing the flow configuration. Thus we assume that the
AP is represented by a solid wall that separates the two
fluids, the stellar and the interstellar wind. Starting from the star,
the stellar wind velocity jumps at the TS. Because it creates an
oblique shock surface, a sonic line can be found at which the
shocked Mach number $M_{2}=1$. Between the stagnation line and the sonic line exists a strong solution ($M_{2}<1$), while beyond the sonic line
the shock solution is weak $M_{2}>1$, that is, the transition from the
supersonic stellar wind remains supersonic. In the tail direction a
triple point $T$ forms, where a slipstream or tangential
discontinuity TD separates the supersonic from the subsonic
flow. At this TD the pressure between the two flows is in
equilibrium. Moreover, from the triple point $T$ a reflected shock
propagates toward the AP, where it is reflected again
toward the TD, where it becomes reflected again. From the
triple point a circular Mach disk extends toward the stagnation
line.
%which has its center at $M$, about 0.6 times the distenace from
%the star to the termination shock at the stagnationm line.

The BS is detached from the blunt AP, where the distance
$\overline{AP\,BS} \equiv \Delta$. Similar to the TS, the BS is an
oblique shock surface, where a sonic line emanates.

To solve the Euler equations the inner boundary values are usually
given by $\rho_{0},v_{0}=v_{s},P_{0}$ (or the temperature $T_{0}$
instead of $P_{0}$). With the knowledge of $R_{TS}$ and the condition
that the fluid inside the TS is polytropic and spherically expanding at
constant speed, we have
\begin{eqnarray}
  \label{eq:ee}
  E &=& \frac{1}{2} \rho v^{2} + \frac{1}{\gamma-1}P \\
  \rho &=& \rho_{0}\frac{r_{0}^{2}}{r^{2}} 
.\end{eqnarray}
Furthermore, the stellar wind at the inner integration boundary is
always supersonic, that is,\ $M_{s,0}>1$. The Mach number $M_{s}$
increases as $r^{2/3}$, meaning that it is by definition inversely proportional
to the square root of the temperature. Thus, we expect huge Mach
numbers at the TS, and following from this, that in this realm the flow is 
``hypersonic'', that is, $M \gg 1$. In turn, the thermal
pressure upstream of the TS is almost zero because it is lower than the ram
pressure at the inner boundary and decreases with $r^{-10/3}$. The
supersonic velocity $v_{0}=v_{TS,1}$ remains constant until it reaches
the TS.

\subsection{ Pressure and temperature in the hypersonic approximation}

We can calculate the pressure downstream of the TS knowing $R_{TS}$
using the jump condition in the momentum equation. Assuming that
the pressure inside the TS is negligible compared to the ram pressure,
we obtain from the Rankine Hugoniot relation for the total momentum density
\begin{eqnarray}
  \label{eq:e3a}
  \rho_{TS,1} v_{s,1,n}^{2} &=&\rho_{TS,2} v_{s,2,n}^{2}  + p_{TS,2} \\
                         &=& \frac{\gamma -1}{\gamma + 1} \rho_{TS,1} v_{s,1,n}^{2}  + p_{TS,2}
,\end{eqnarray}
with $\rho = \rho_{0} r_{0}^{2} /r^{2}$ , and $r_{0}$ is a reference
radius ($r_{0}=1$\,AU for the heliosphere and
$r_{0}=0.03$\,pc for \LamC).  We easily derive
the shocked pressure $P_{TS,2}$ as
\begin{eqnarray}
  P_{TS,2} = \frac{2}{\gamma+1} \frac{r_{0}^{2}}{R_{TS}^{2}}\rho_{0} v_{0}^{2} \sin^{2}\vartheta
,\end{eqnarray}
where $\vartheta$ is the \SA, which is the angle between the upstream
velocity and the shock.

At the nose $\vartheta=\pi/2$ and all the values on the right-hand side
are known by assumption.  We can furthermore assume that the pressure in
the region given by the intersection of the sonic line, the TS, the AP,
and the stagnation line, is almost constant, because it is a subsonic
flow. Thus we obtain a relation between the stellar-centric distances
along the TS and the \SA, which is not identical with a
stellar-centric angle $\Phi$ to the same point on the TS.

Another nice feature of the hypersonic flow is that the temperature
directly beyond the TS depends only on the initial
velocity and the \SA because the Mach number
$M_{1}^{2}$ increases inversely proportional to the temperature
$T_{1}=T_{0}(r_{0}/r)^{-4/3}$. With
this,  we find
\begin{eqnarray}
  \label{eq:t1}
 T_{2} &=&  \frac{2\gamma (\gamma - 1) \Ms }{(\gamma + 1)^{2} } T_{1} =
 \frac{2 (\gamma - 1) m
     v_{S,1}^{2}\sin^{2}\vartheta }{(\gamma + 1)^{2}  k T_{1}
 } T_{1} \\
&=& \frac{2  (\gamma - 1) }{(\gamma + 1)^{2} } \frac{m_{p}}{k}
v_{0}^{2} \sin^{2}\vartheta
,\end{eqnarray}
where $m$ is the mass of a fluid particle and $k$ is the Boltzmann
constant. Here we  additionally assumed that $m=m_{p}$ (proton mass $m_{p}$).

This has the nice consequence that the temperature
downstream of the TS determines the \SA $\vartheta$ when $v_{0}$ is
known. Furthermore, $v_{0}$ can be determined measuring $T_{TS}$, where
$\vartheta = \pi/2$. For \LamC the temperature at the TS
for $\vartheta=\pi/2$ and $v_{0}=2500$\,km/s is
$T_{2,TS} = 1.2\cdot10^{8}$\,K. From an observational point of view, the BS
is easier to observe, and for $v_{0}=80$\,km/s the temperature
is $T_{2,BS}= 1.4\cdot10^{5}$\,K.

Typically, $v_{0}$ is known as the terminal velocity from stellar
wind models,  therefore by measuring the temperature at the TS in
the nose direction (i.e., at the stagnation line), we can determine if the
polytropic coefficient  $\gamma= 5/3$.\ This amounts to a test whether the assumption
of a mono-atomic gas holds. In addition, the same is true
for the BS when the interstellar Mach number is high
enough.

This holds true for most of the models where no other interaction with
the hypersonic stellar wind appears, like charge exchange with
neutrals. This includes cooling or heating inside the TS because
there the temperatures and densities are low, so that neither
cooling
nor heating is effective (see below). This should also apply to MHD
models in which the magnetic field pressure inside the TS is
negligible compared to the ram pressure.  Models including mass,
momentum, and energy loading have to be excluded because the
stellar wind speed is affected in those.

\subsection{Termination shock distance}

Next we determine the TS distance. First we recognize
that since there is no mass flow through the AP,
$v_{n,I,2}=v_{n,S,2}=0$, where $I \text{ and }S$ denote the interstellar and
stellar conditions, and $n$ describes the velocity
component normal to the shock. 

Inserting this into the momentum conservation, we derive
at the AP $P_{I,2}(AP)=P_{S,2}(AP)$. Because the total momentum is
conserved along a flow line, which here is the stagnation line, the
pressure at the AP is the same as that directly after the BS:
\begin{eqnarray}\label{bs}
  P_{I,2}(AP) &=&\rho_{I,2}v^{2}_{I,2}+P_{I,2}(BS)\\\nonumber
              &=&\rho_{I,1}v^{2}_{I,1}+P_{I,1}(BS). 
\end{eqnarray}

Inside the AP a similar condition holds for the stellar
wind (replacing the index $I$ by $S$ and BS by TS). Thus we derive at the
AP
\begin{eqnarray}
  \label{ap}
  \rho_{I,1}v^{2}_{I,1}+P_{I,1}(BS) = P_{I,2}(AP) = P_{S,2}(AP)
  \\\nonumber
= \rho_{S,1}v^{2}_{S,1}+P_{S,1}(TS).
\end{eqnarray}

We can express the density of the stellar wind by its mass-loss rate
$\dot{M}_{\star}$ and its terminal velocity
$||\vec{v}_{\infty}||=||\vec{v}_{0}||$ ($=v_{S,1}=\const$ in the
supersonic stellar wind region) by
\begin{eqnarray}
  \label{eq:mdot}
  \dot{M_{\star}} = 4\pi r^{2}\rho_{S,1}v_{0}
\end{eqnarray}
or 
\begin{eqnarray}
  \label{eq:m2}
  \rho_{S,1}=\frac{\dot{M}_{\star}}{4\pi r^{2}v_{S,1}}
.\end{eqnarray}
Taking Eq.~(\ref{ap}), rearranging it, and inserting Eq.~(\ref{eq:m2}),
leads to
\begin{eqnarray}
  \label{aps}
  \frac{R^{2}_{TS}}{r_{0}^{2}} &=&
  \frac{\rho_{0}v_{0}^{2}}{\rho_{I,1}v_{I,1}^{2}+P_{I,1}(BS)-P_{S,1}(TS)}
  \\\nonumber
&=& \frac{\dot{M_{\star}}v_{0}}{4\pi\left(\rho_{I,1}v^{2}_{I,1}+P_{I,1}(BS)-P_{S,1}(TS)\right)} 
\end{eqnarray}
with $\rho_{S,1}=\rho_{0} r_{0}^{2}/r^{2}$.

The thermal pressures in the supersonic media can usually be neglected
compared to the ram pressure, at least for high Mach number
flows. Then Eq.~(\ref{aps}) reduces to
\begin{eqnarray}
  \label{aps2}
  \frac{R^{2}_{TS}}{r_{0}^{2}} = \frac{\rho_{0}v_{0}^{2}}{\rho_{I,1}v_{I,1}^{2}} = \frac{\dot{M_{\star}}v_{0}}{4\pi\rho_{I,1}v^{2}_{I,1}} 
.\end{eqnarray}

$R_{TS}$ is the TS distance at the stagnation line. We wish to
emphasize that the TS distance is the only distance that can be
determined without any other assumption in addition to the momentum
conservation and the polytropic behavior of the stellar wind. The TS
distance is not the stagnation distance. The latter is
defined as the distance at the stagnation line where the (shocked)
interstellar flow speed and the shocked stellar wind speed are both
zero, which is the AP distance $R_{AP}$.

The $r$-dependence of the density and velocity is unknown in the shocked
stellar wind region, that is,\ in the inner astrosheath. Similarly
unknown is
the functional dependence of those in the shocked interstellar wind
region, that is, in\ the outer astrosheath. Therefore the stagnation distance or
AP standoff distance cannot be determined without further
assumptions.

\subsubsection{Astropause distance I: in upwind direction}

The AP or stagnation distance is often described using a flow
potential \citep{Parker-1961, Suess-Nerney-1990} or using the thin-shell approximation
\citep{Baranov-etal-1971,Dyson-1975,Raga-etal-2014}, where a
dependence of $r$ from  the stellar-centric angle $\Phi$ to a
point on the AP or BS is formulated. As was discussed in
\citet{Zank-1999}, these are a rather poor approximations.

\begin{figure}[t!]
  \centering
  \includegraphics[width=0.95\columnwidth]{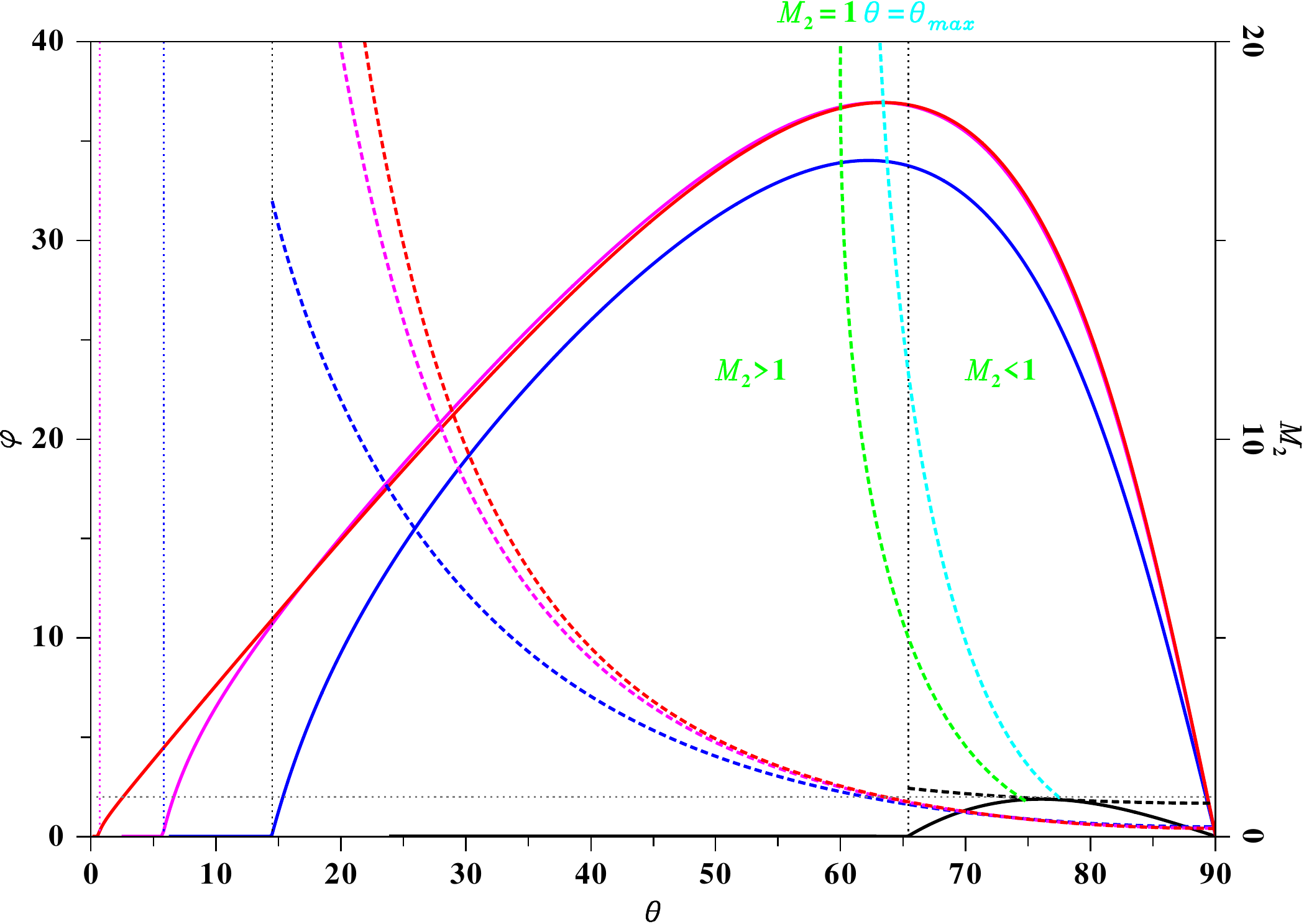}
  \caption{ Dependence of the \FA with respect to the \SA for
    different Mach numbers $M_1$ (solid lines). The  black,
    blue, magenta, and red lines correspond to $M_{1}=1.1,4,10
,\text{and }100$,
    respectively. To the right of the dotted vertical lines a physical
    solution exists for the corresponding Mach number, while to the
    left it does not. The dashed lines show the dependence of the
    shocked Mach number $M_{2}$ from the \SA $\vartheta$ (scale on the
    right vertical axis). If $M_{1}$ varies, this
    leads not to strong variations of $\vartheta$ around the sonic
    points, where $M_{2}=1$ and $\vartheta \approx 57-64$ degrees for
    all Mach numbers $M_{1}$. But the variation of the \FA $\varphi$
    (left vertical axis) is large, as shown by the green dashed
    line. This line also separates the region between the weak
    solution ($M_{2}>1$, left from the green dashed line) and the
    strong solutions ($M_{2}<1$ right of the green dashed line). At this
    green dashed line $M_{2}=1$ is close to the cyan dashed
    line, where the \SA ($\vartheta$) reaches maximum.}
\label{rh:3}
\end{figure}

The reason is that the shock normal is in general not parallel to the
radius vector (see Fig~\ref{rh:3}).  The \SAs $\vartheta_{sl,i}$ and
the \FA $\varphi_{sl,i}$ at the sonic points, which are the points at
the intersection of the sonic line with TS (index S) and with the BS
(index I), see Fig.~\ref{rh:1}, can be calculated setting $M_{2} = 1$
to determine the \FA. For the interstellar flow this is not a problem
because there we know the Mach number $M_{I,1}$ in front of the BS and
for the stellar wind flow we assume $M_{S,1}\gg 1$.  The green line shows that the \FA $\varphi$ changes dramatically when the \SA
$\vartheta$ varies only very slightly. For a wide range of \SAs the
flow pattern can therefore be different, but for Mach numbers $M_{I}>2,$ the
\FAs $\varphi$ and the \SA $\vartheta$ are approximately constant. The
overall shock structure depends on both the stellar and interstellar
ram pressure, but the two angles $\varphi$ and $\vartheta$ do not, which means that
it cannot be expected that purely geometrical considerations lead to
reliable results.

% For the stellar wind at the termination shock, we
% assume that $M_{1,s}>>1$ and thus we neglect the $r$-dependence for
% that approach. The caluclations then yield for \LamC with
% $M_{1,s} \approx 900$ and $M_{1,I} \approx 10$
% \begin{eqnarray}\nonumber
%   \vartheta_{sl,s} &\approx& 63^{o} \qquad \varphi_{sl,s} = 37^{o}    \\\nonumber
%   \vartheta_{sl,I} &\approx& 63^{o} \qquad \varphi_{sl,I} = 36^{o}    \\\nonumber
%   \vartheta_{sl,I} &\approx& 63^{o} \qquad \varphi_{sl,I} = 24^{o}
%   \qquad \mathrm{for} \tilde{M}_{ism}= \frac{1}{4} M_{ism}  \\\nonumber
% \end{eqnarray}

The deviation of the shocked interstellar flow in particular
violates
the assumption of a parallel subsonic flow at infinity and the
irrational condition that is needed to derive the stagnation
distance from the flow potential. The thin-shell approximation is usually
violated because the distance between the TS and AP, and to a lesser extent
between the AP and BS, are not shorter than the TS distance. As we
 discuss below, the distance between the AP and the BS shrinks
remarkably as a result of cooling effects. But then the structure of the
astrosphere is more complicated.

There is no simple functional dependence between the stellar-centric
angle $\Phi$ and the ISM \SA $\vartheta_{ism}$ or the ISM \FA
$\varphi_{ism}$. The latter depends on the Mach number $M_{ism}$ of
the inflowing ISM and also on the shape of the
BS. The latter shape is also influenced by the AP, which in turn is
formed by the shape of the TS. The curvature of the latter is also
shaped by the ram pressure of the ISM (Mach number $M_{ism}$),
which is shown by the following consideration: With a varying ram pressure
of the ISM, the distance of the TS changes. If we assume that the
region between the TS at the stagnation line and sonic point on the
TS can be described by a sphere with radius $R_{TS}$, then it is
obvious that the curvature of the TS around $R_{TS}$ changes as
$R_{TS}^{-1}$.  Then for low decreasing interstellar ram pressures
(Mach number $M_{ism}>1$) the \FA $\varphi_{sl,ism}$ at the sonic
point decreases and the upwind distance $\Delta$ between AP and BS increases,
while for high increasing interstellar Mach numbers $M_{ism}\gg1$ the
distance $\Delta$ decreases, but the sonic \SA $\vartheta_{sl,ism}$
and the sonic \FA $\varphi_{sl,ism}$ remain approximately
constant. Thus the stellar-centric position angle $\Phi$ must depend
on $\vartheta_{ism}$ and $\varphi_{ism}$. The sonic points on the TS
are given by the stellar Mach number $M_{s}$ , which is always huge, and
thus we can conclude that the sonic \SA $\vartheta_{sl,s}$ and the
sonic \FA $\varphi_{sl,s}$ are not affected by the change of $M_{ism}$
, only the curvature of the TS is changed and moves the sonic lines
toward or away from the inflow axis.

The shape of the AP cannot be determined easily because this
is a pressure equilibrium surface, specifically, a tangential
discontinuity. Across this structure only the thermal pressure has to
be balanced on both sides, while the tangential velocity, the
temperature, and the density may jump. The normal velocity component is
zero because no mass is transported through a tangential
discontinuity. We just state here that a contact discontinuity
is a different feature, which appears in an MHD shock structure, where
the normal magnetic field does not vanish. Then the density can
jump, while all other quantities (tangential velocity and magnetic field, the
sum of thermal and magnetic field pressure) are continuous across a
contact discontinuity. Nevertheless, the determination of the
AP shape can be formulated as an inversion problem, knowing
the (parabolic) shape of the BS, the shape of the AP, especially the
distance between AP and BS, can be estimated
\citep{Schneider-1968,Schulreich-Breitschwerdt-2011}.

The stellar-centric AP distance can also be calculated using the
approach to determine the distance between the AP and the BS, as we
discuss below.

\subsubsection{Bow shock distance to the astropause}
 We assume that the AP is locally approximated by a solid sphere with radius
 $R_{AP}$. For this blunt body  an approximate BS distance  using
   the two-dimensional continuity and energy equation can be
determined \citep{Olivier-2000} for $\gamma=5/3$:
\begin{eqnarray}
  \label{eq:o1}
  \Delta_{BS} = 2 \hat{\Delta} \frac{\rho_{I,1}}{\rho_{I,2}} R_{AP} = 2
  \hat{\Delta} \frac{1}{s} R_{AP} = \tilde{\Delta} R_{AP}
,\end{eqnarray}
where $\Delta_{BS}$ is the distance between the stellar-centric BS
and the AP ($\Delta_{BS}=R_{BS}-R_{AP}$), $s$ is the
compression ratio $s=\rho_{I,2}/\rho_{I,1}$ and
$\tilde{\Delta}=2\hat{\Delta}/s$ is given by
\begin{align}
  \label{eq:o2}
& \tilde{\Delta} 
% = \frac{2}{s}
%   \dfrac{s\left[\sqrt{\dfrac{1}{4}(1+a)^{2}-\dfrac{1}{3s}(1+2a)}
%       -\dfrac{1}{2}(1+a)\right]+1}{\dfrac{4}{3}+\dfrac{2}{3}a-\dfrac{2}{s}}\\\nonumber
% &&
= \dfrac{3}{2s+3a\,s-3}\left(\sqrt{\dfrac{s^{2}}{4}(1+2a)^{2}-\dfrac{s}{3}(1+2a)}-\dfrac{s}{2}(1+a)+1\right) \\
&\tilde{\Delta}(a=0)=  \dfrac{3}{2s-3}\left(\sqrt{\dfrac{s^{2}}{4}-\dfrac{s}{3}}
      -\dfrac{s}{2}+1\right) \\
&\tilde{\Delta}(a=1)=  \dfrac{3}{5s-3}\left(\sqrt{\dfrac{9s^{2}}{4}-s}-s+1\right)
\end{align}
where $a$ is the derivative of the normalized tangential velocity at
$R_{AP}$, that is,\ $a=\partial{u}/\partial{\varphi}$. Assuming that the tangential
velocity vanishes at the BS in the inflow direction, we obtain
$\tilde{\Delta} \approx 0.44$ for the above compression ratio
$s=3.87$ for the ISM around \LamC. 

\begin{figure}[t!]
  \centering
  \includegraphics[width=0.9\columnwidth]{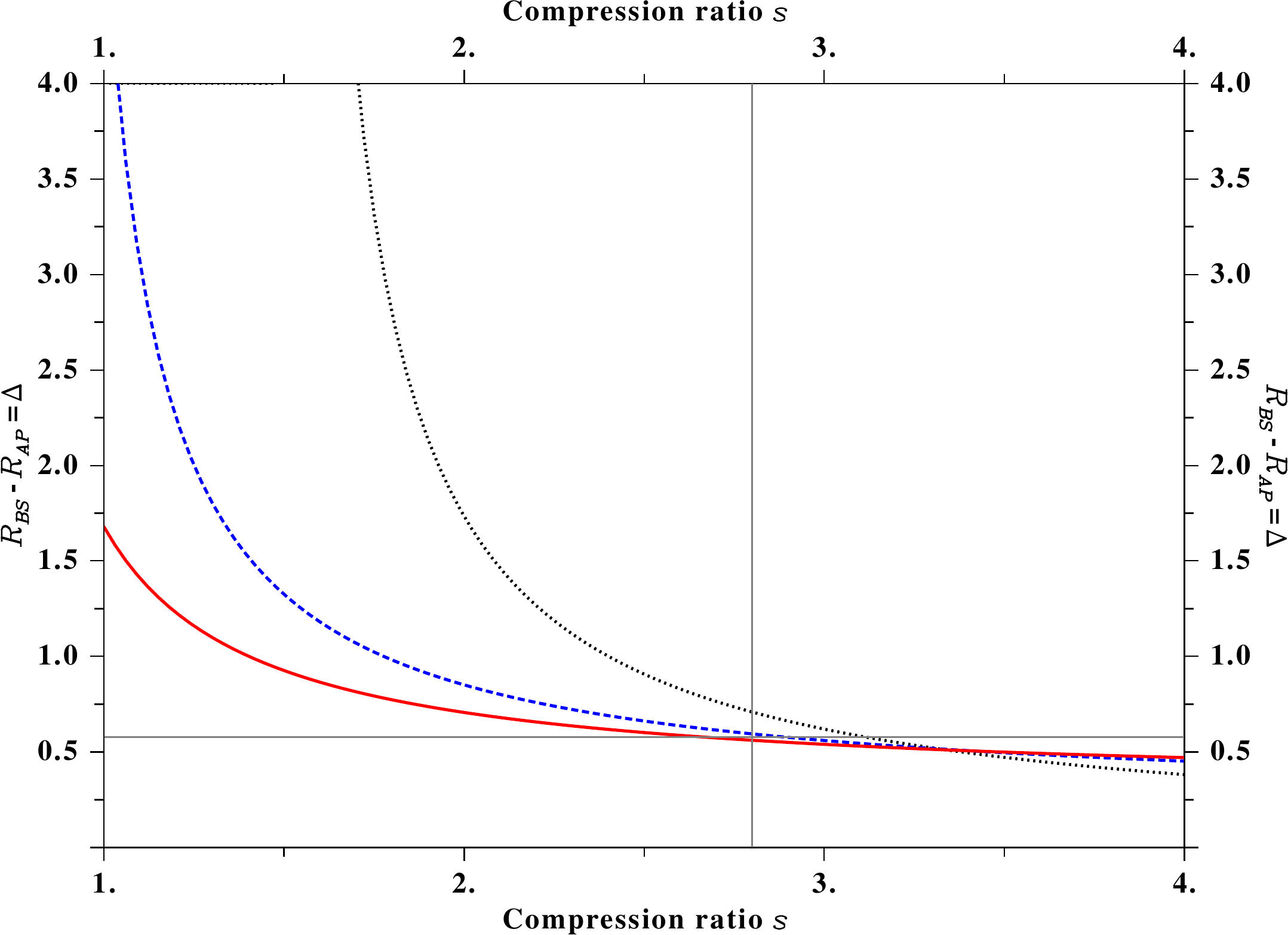}
  \caption{Distance between BS and AP as a function
    of the compression ratio $s$. The red line shows the $\Delta(s)$
    with $a = 1$, the blue dashed line the same for $a=0.5$, and the black
    dashed line for $a=1$. }
  \label{fig:11}
\end{figure}

Earlier approaches \citep[e.g.,\ ][]{Barnette-1993}  found numerically
\begin{eqnarray}
  \label{eq:il}
  \Delta_{BS} = 0.855 \frac{\rho_{ISM}}{\rho_{s}} R_{AP} = 0.855 \frac{R_{AP}}{s}
.\end{eqnarray}
For non-reactive gases 
\citet{van-Dyke-1958}  determined $\Delta_{BS}$ also numerically  by
\begin{eqnarray}
  \label{eq:dl}
  \Delta_{BS} = 0.41 \frac{\rho_{ISM}}{\rho_{s}} R_{AP} = 0.41 \frac{R_{AP}}{s}
.\end{eqnarray}
The flow pattern around a sphere with diameter $R_{TS}$ was calculated
and experimentally verified by the above authors. If we assume that
around the nose the AP can be described by a sphere with the radius
$R_{AP}$, the stellar-centric distance of the AP, we can use the
above estimates to determine $\Delta$. The BS distance from the
center is
\begin{eqnarray}
  \label{eq:bs}
  R_{BS} = (1 + \Delta_{BS}) R_{AP}
.\end{eqnarray}

\citet{Farris-Russell-1994} approximated the blunt body by a conic
section, and with a given radius of curvature, these authors found the
bow shock distance as
\begin{eqnarray}
  \label{eq:fr}
  R_{BS} &=& R_{C} \left(\frac{R_{AP}}{R_{C}} + 0.8 \frac{(\gamma -1)
      M_{1}^{2}+2}{(\gamma+1)(M_{1}^{2}-1)} \right)  \\\nonumber
        &=& R_{C} \left(\frac{R_{AP}}{R_{C}} +  \frac{1.6}{(s - 1)(
            \gamma +1)} \right)
\end{eqnarray}
which can also be applied to low Mach numbers, for which the BS
then moves to infinity. $R_{C}$ is the radius of curvature defined as
$1/\kappa,$ where $\kappa$ is the curvature. For a sphere with radius
$R_{AP}$ , we have $R_{C}=R_{AP}$. The latter equality in
Eq.~(\ref{eq:fr}) follows from the Rankine-Hugoniot equations for the
compression ratio. 

For hypersonic flows Eq.~(\ref{eq:fr}) can be simplified to
\begin{eqnarray}
\label{eq:fr2}
   R_{BS} = R_{C} \left(\frac{R_{AP}}{R_{C}} + 0.8 \frac{\gamma
       -1}{\gamma+1} \right) =  R_{AP}\left(1 + \frac{0.8}{s}
   \right) = 1.2 R_{AP} 
.\end{eqnarray}
Equations~(\ref{eq:o2}) and~(\ref{eq:fr}) give almost identical results for
the respective parameters, and the hypersonic limit Eq.~(\ref{eq:fr2}) is
also acceptable, especially for \LamC, as can be seen in Table~(\ref{tab:c3}). 

The problem of determining $R_{AP}$ remains, however.

\subsubsection{Astropause distance II}

We use the same approach as above to determine the AP
distance from that of the TS. Assuming now that the supersonic wind
approaches a hollow sphere, we obtain with the compression ratio
$s_{s} \approx 4$ of the stellar wind $\tilde{\Delta} \approx 0.38$.
This approach also nicely reproduces the AP distances
$R_{AP}= (1+\Delta)R_{TS}$ , as given in Table~\ref{tab:c3}.

\subsubsection{Tail termination shock distance}

In the tail direction, the flow along the AP and the
tangential discontinuity is almost parallel. The only pressure
that has to be balanced, therefore, is the thermal pressure of the interstellar
medium, which then also has to be balanced in the subsonic tail
region. By replacing the ram pressure in Eq.~(\ref{aps}) by the
thermal pressure, the tail distance of the TS can be determined as
\begin{eqnarray}
\label{eq:tail}
   R_{TS}(tail) & = r_{0} \sqrt{\dfrac{\rho_{S,1}v_{S,1}^{2}}{P_{I,1}}}
.\end{eqnarray}

This is not a good approximation, see Table~\ref{tab:c2}, because
instead of the shocked, the interstellar thermal pressure should be used, which can be higher than the undisturbed
pressure. Nevertheless, the undisturbed pressure alone can
be determined analytically.

\subsubsection{Temperature and pressure}
\label{sec:p1}
For high Mach number flows, we can replace the Mach
number by $M_{1}^{2} = \mu v_{1}^{2}/(\gamma k T _{1})$, from which follows
\begin{eqnarray}
  \label{eq:t1a}
  T_{2} = \frac{2 (\gamma - 1)}{(\gamma + 1)^{2}}
  \frac{\mu}{k}\sin^{2}\vartheta v_{1}^{2} 
,\end{eqnarray}
with the average mass $\mu = \sum_{s}a_{s}m_{s}$ and the Boltzmann
constant $k$. $a_{s}$ and $m_{s}$ give the abundance and mass of a
particle of species $s \in\{\p, \He^{+}, \He^{++}, ....\}$. We also dropped the indices $S\text{ and }I$ because Eq.~(\ref{eq:t1})
holds in general.

Hence, the shocked temperatures upwind and downwind of the TS are
identical, and they vary like $\sin^{2}\vartheta$
along the TS. This means that
in the region between the nose
($\vartheta = 90^{o}, \sin^{2}\vartheta = 1$) and the sonic point
at the BS ($\vartheta \approx 63^{\deg}\ \mathrm{and} \sin^{2}\vartheta \approx 0.7$), the
temperature directly beyond the TS changes by 30\%. The density jump
for hypersonic flows is roughly 4, and consequently, the thermal pressure varies by 30\% along the TS in the subsonic region.

Moreover, if we  know the temperature along
the shock, the \SA $\vartheta$ can be determined for hypersonic
flows, which then determines
the flow pattern directly behind the shock.  This is also valid for
interstellar flows with sufficiently high Mach numbers, especially
around the nose region of the BS, where $\vartheta > 60^{\circ}$. This means that the temperature in the nose direction can be
used to determine the
interstellar flow speed using Eq.~(\ref{eq:t1}).

Between the sonic lines the flow is subsonic, therefore the thermal
pressure is nearly constant. This is also true beyond the Mach disk
(see Sect. 4.3.6), and the pressure can therefore be estimated by calculating
the pressure at the nose for the interstellar and stellar fluid, and the thermal pressure at the tail for the stellar
flow. For the hypersonic case we find from the Rankine-Hugoniot
relation
\begin{eqnarray}
  \label{eq:p1}
  P_{2} = \frac{2 \sin^{2}\vartheta}{\gamma + 1} \rho_{1} v_{1}^{2}
.\end{eqnarray}

Applying these approximations to the stellar wind, which is highly
supersonic (for the $\lambda$ Cephei stellar wind, Mach numbers at the
TS are $M_{1,S}\approx 7000$), the stellar wind thermal pressure is
negligible compared to the stellar wind ram pressure and thus does not
influence the dynamics of the flow inside the TS. This is convenient,
because only the stellar wind density and velocity can be determined
from observations and the theory of stellar atmospheres. But that is
all what is needed to describe the dynamics of the flow inside the TS
in the \sfl model.

\subsubsection{Mach disk}

The numerical experiments show that the Mach disk can be
fitted by a circle, the center of which is shifted  from the star
toward the tail. The translation parameter and the radius of the
circle have not been determined analytically up to now.

\section{Hydrodynamic shocks: Including cooling and heating
  functions}\label{cool}

The great extent of astrospheric shocks around hot stars can cause an
efficient cooling of the flow inside the BS. Many
different cooling functions are discussed in the literature
\citep[see][]{Rosner-etal-1978,Dalgarno-McCray-1972,Sutherland-Dopita-1993,
  Schure-etal-2009,
  Mellema-Lundqvist-2002,Townsend-2009,Reitberger-etal-2014a},
which reflects the different compositions and states of the ISM.  The
piecewise analytic representation by \cite{Siewert-etal-2004} lies in
between these extremes, and we have chosen this representation for our
model of \LamC. The cooling functions can differ by orders of
magnitude, especially at lower temperatures
\citep{Dalgarno-McCray-1972}, depending on the metallic
abundances. These lower temperatures are usually reached when a low
Mach number BS appears, which is typically the case for low, but still
supersonic interstellar wind velocities. In the tail regions such
lower temperatures can also be reached. The shocked stellar wind is usually
above $10^{8}$\,K, except for the flanks of the astrosphere, where
oblique shock transitions with weak solutions appear ($M_{2}>1$).

The thermal pressure is higher than the ram
pressure in subsonic regions. By neglecting the ram pressure in the energy equation and
assuming an polytropic behavior of the gas, we can therefore write the energy
equation approximately as (also discussed in
\citet{Scherer-etal-2014})
 \begin{eqnarray}
  \label{eq:c1}
\div \left(\frac{\gamma}{\gamma -1}P + \frac{1}{2} n
  m_{p} v^{2}  \right) \vec{v} &\approx&
\frac{\gamma P v}{(\gamma-1)L_{cool,s}} = \frac{5 n k T v}
{L_{cool,s}};\\\nonumber
 &=& -n^{2} \Lambda(T)
\end{eqnarray}
replacing $\nabla$ by the inverse subsonic
cooling length,  $L_{cool,s}$ can be
estimated as
\begin{eqnarray}
  \label{eq:c2}
  L_{cool,s} \approx \dfrac{\gamma P v}{(\gamma-1) n^{2} \Lambda(T)} =
  \dfrac{5 k T v}{n\Lambda(T)}
,\end{eqnarray}
and the subsonic  cooling time $\tau_{cool,s}$ 
\begin{eqnarray}
  \label{eq:c3}
  \tau_{cool,s} = \dfrac{L_{cool,s}}{v} = \dfrac{5 k T }{n\Lambda(T)}
.\end{eqnarray}
This cooling time is by up to a factor 3 the same as that given in
\citet{Sutherland-Dopita-1993}. 
Vice versa, in a hypersonic wind we neglect the thermal pressure and
obtain analogously
\begin{eqnarray}
  \label{eq:c5}
  L_{cool,h} \approx \dfrac{m_{p} v^{3}}{2 n \Lambda(T)} \qquad
  \mathrm{and} \qquad\tau_{cool,h}
  =  \dfrac{m_{p} v^{2}}{2 n\Lambda(T)}
,\end{eqnarray}
where $L_{cool,h}$ and $\tau_{cool,h}$ are the hypersonic cooling
length and time. 

The heating function by photoionization and some supplemental
heat source is discussed in \citet{Kosinski-Hanasz-2006}, but see
also \citet{Reynolds-etal-1999}:
\begin{eqnarray}
  \label{eq:h1}
  \Gamma = n^{2} G_{0} + n G_{1} 
.\end{eqnarray}
With the constants $G_{0} = 10^{-24}$\,erg cm$^{3}$ s$^{-1}$ and
$G_{1} = 10^{-25}$\,erg\,s$^{-1}$ and replacing the right side of
Eq.~(\ref{eq:c1}) by $\Gamma$, we obtain
\begin{eqnarray}
  \label{eq:h2}
  L_{heat,s} = \dfrac{5 k T  v}{n G_{0} + G_{1}} \qquad \mathrm{and} \qquad
  L_{heat,h} = \dfrac{m_{p} v^{3}}{2 (n G_{0} + G_{1})} 
,\end{eqnarray}
and the heating times $\tau_{heat,s}, \tau_{heat,h}$ by dividing the
respective heating lengths by the speed $v$.

The cooling and heating lengths can be interpreted as the distances at
which the flow is significantly cooled or heated. Moreover, it gives an
$r$-dependence of the flow in the outer astrosheath, which allows
determining the distance between the BS and the AP, which is in the
order of a cooling length (depending on the shocked temperature).

\begin{figure}[t!]
\label{fig:cool2}
  \centering
  \includegraphics[width=0.9\columnwidth]{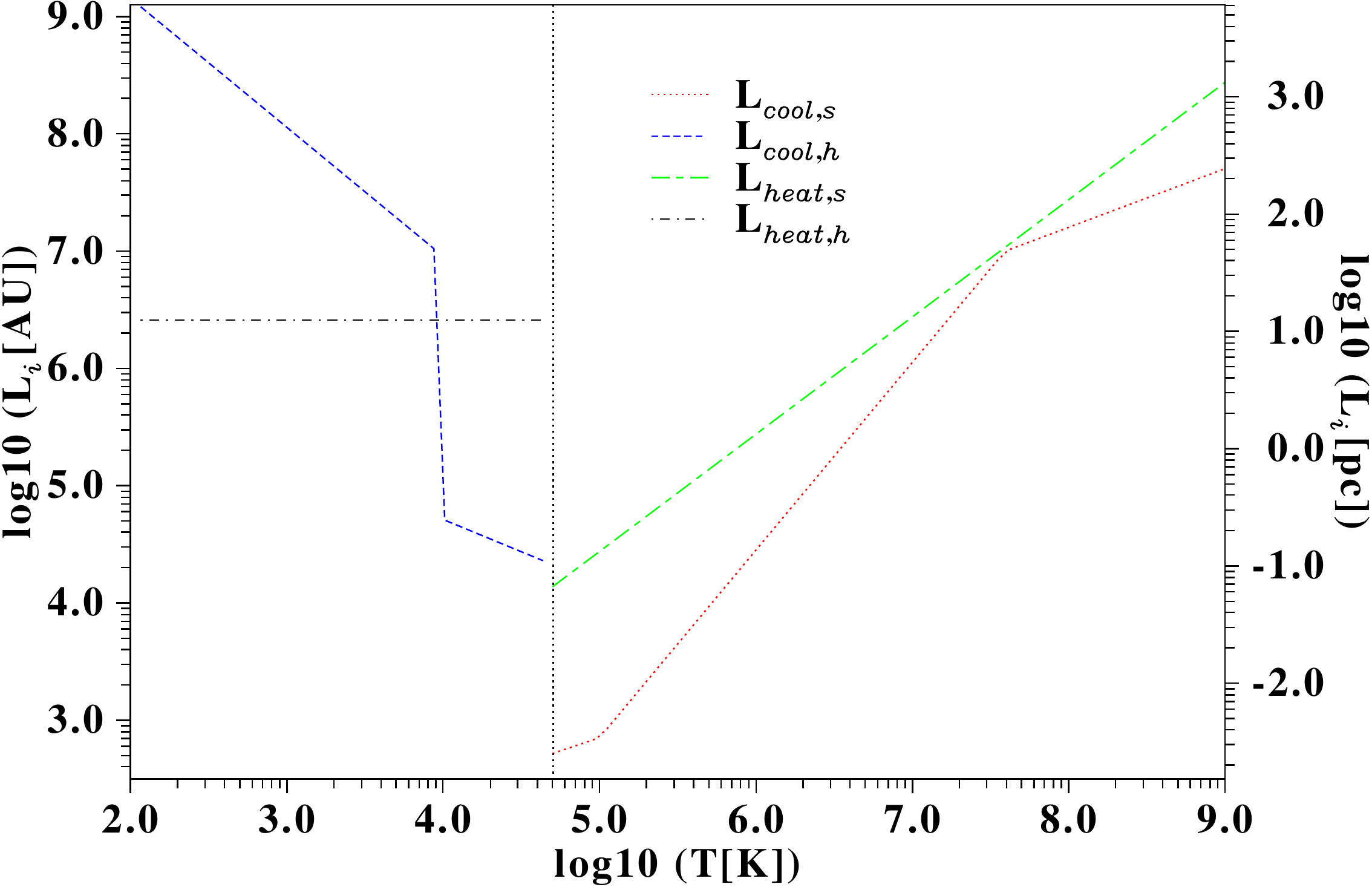}
  \caption{Cooling lengths. The number density and speed are fixed at $n=
     10$\,cm$^{-3}$ and $v=80$\,km/s. All scales are
   logarithmic. On the x-axis we display the
temperature, while the characteristic length is shown in AU on
the left y-axis and
 in parsec on the right y-axis. The dashed vertical line denotes the
 temperature where the ram pressure equals the thermal pressure.}
\end{figure}

\begin{figure}[t!]
\label{fig:cool3}
  \centering
  \includegraphics[width=0.9\columnwidth]{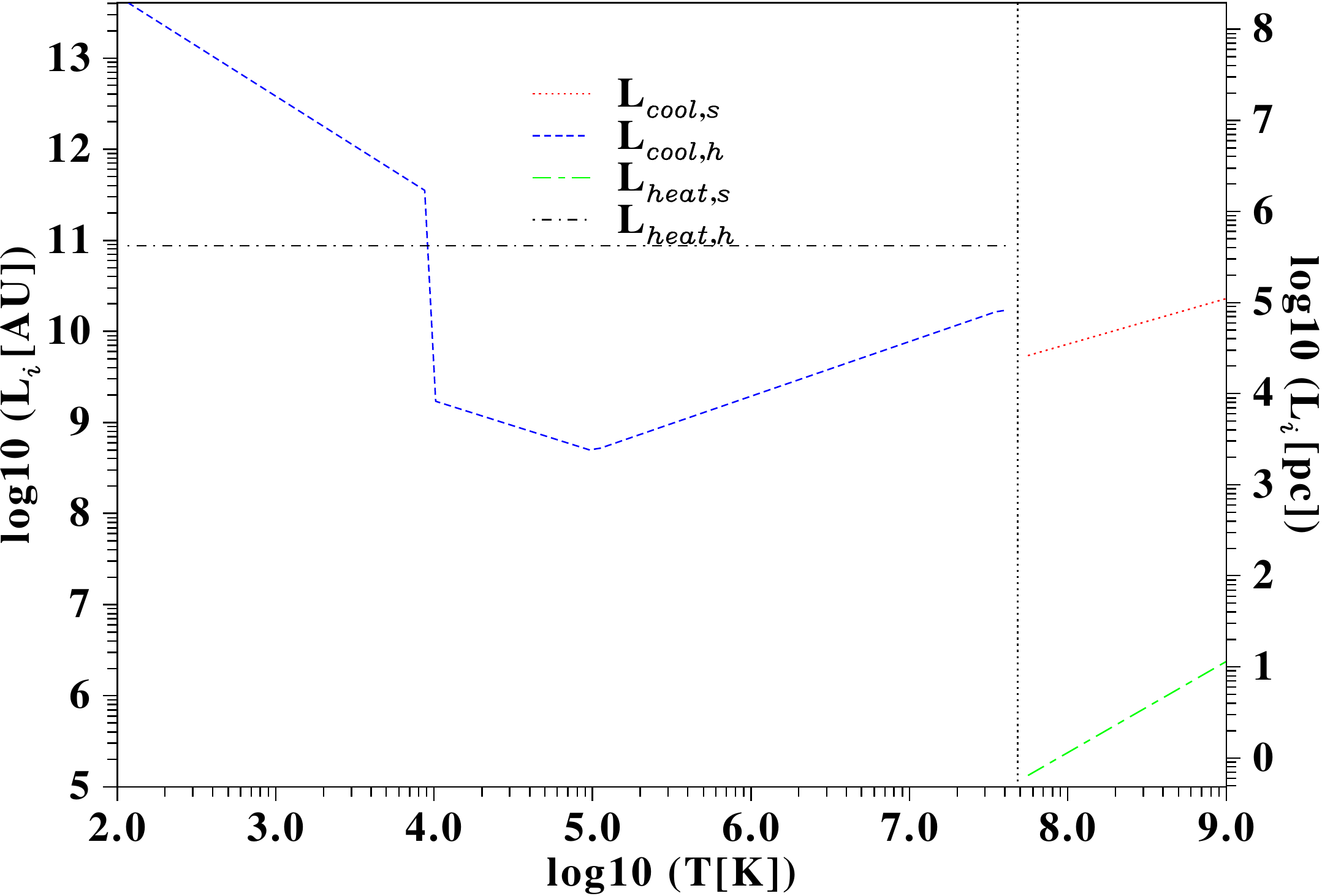}
  \caption{Cooling lengths. The number density and speed are fixed at $n=
     10$\,cm$^{-3}$ and $v=2500$\,km/s. For the meaning of the
     subscripts see text.}
\end{figure}

Even without running numerical models, we can elicit some information about
the thickness of the outer astrosheath, especially about the distance
between the BS and the AP \citep[see][]{Scherer-etal-2014}. For
convenience these estimates are repeated  here.

The hot shocked ISM is mainly affected because the number
densities in the shocked ISM are higher by a factor of 3.6, which is
the compression ratio between the shocked and unshocked ISM, and
according to Eq.~(\ref{eq:t1}), the shocked temperature
$T_{2,I} \approx 8.\cdot10^{4}$\,K. In the range where
$T\approx 10^{3}$--$10^{5}$\,K, most of the discussed cooling functions
are similar, which means that they jump
three orders of magnitude around $10^{4}$\,K. Above that temperature,
the different cooling functions lead to a more or less efficient
cooling until the temperature falls below $\approx 10^{4}$\,K. Thus,
the time in which the bow shock reaches its final position can differ
because of the cooling efficiency in the outer astrosheath. Moreover,
because the astropause is a pressure equilibrium surface that
balances the shocked stellar wind and ISM thermal pressure, it follows
from the ideal gas law $P=2nkT$ that the number density $n$ in the
outer astrosheath toward the AP must increase to balance the stellar
wind thermal pressure. This is different from a purely hydrodynamic
model, where the density and pressure are almost constant in both
astrosheaths. (The densities differ on both sides of the AP, but the
pressures are balanced.)
\begin{figure*}[t!]
  \centering
  \includegraphics[width=0.445\textwidth]{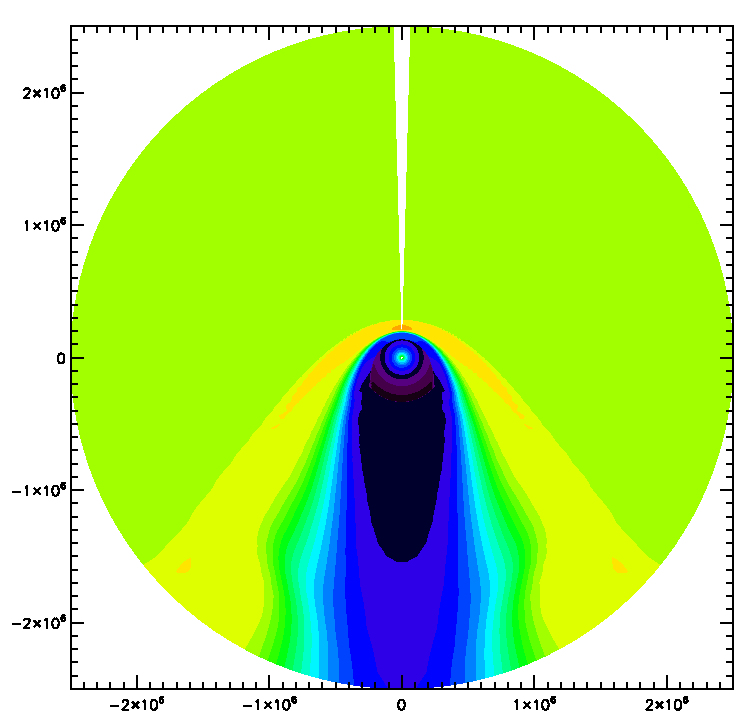}\hfill
  \includegraphics[width=0.5\textwidth]{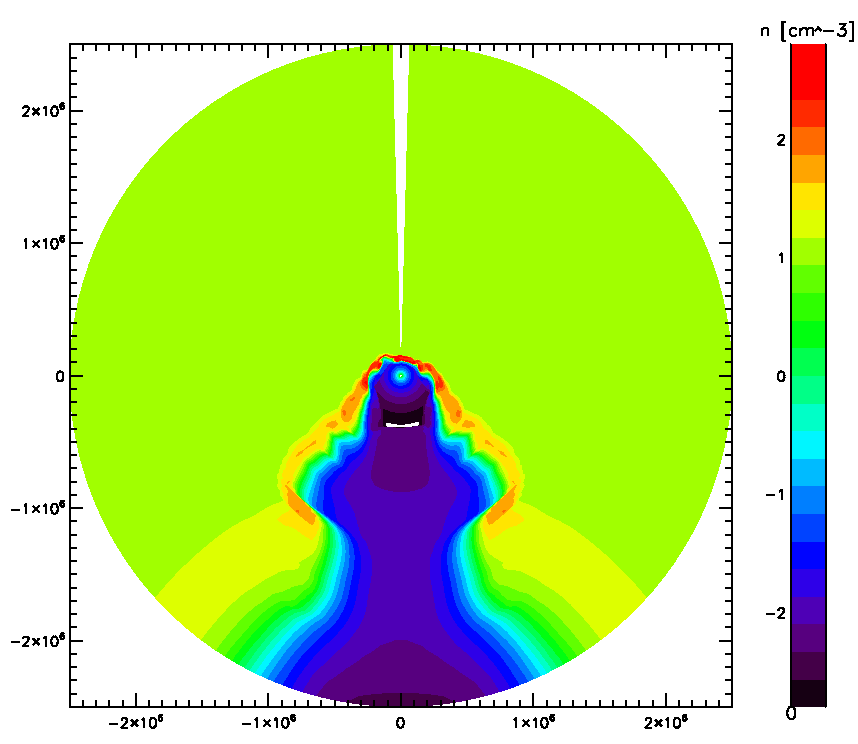}
  \caption{Number density for the pH-model (left panel) and the CH
 -model (right panel). Both models are still not in quasi-stationary
 equilibrium, the physical time is ca. 160\,kyr. The length scale is
 in AU, which corresponds to a box size of $24$\,pc.}
\label{fig:n-ph}
\end{figure*}

In the inner astrosheath, the region between the AP and the TS,
the number density is on the order of $n=10^{-3}$\,cm$^{-3}$ and the
velocities are on the order of $v=600$\,km/s, and thus cooling and
heating length scales, depending on $v$ and $n,$ become huge: the
cooling is no longer important in the inner astrosheath and inside
the TS. The heating length scale depends only on the number density
$n$ and also increases to scales much larger than the distances
inside the AP. This holds true at the inner integration boundary, but
inside of this, cooling and heating can become more important because
the number density increases with decreasing stellar-centric
distance. For stellar wind
velocities of $v=2500$\,km/s, it can easily be estimated that the temperature must be higher than
$T\approx 10^{7}$\,K (see Fig.~\ref{fig:cool3}). Figure 5 shows that in the hypersonic region cooling dominates for
temperatures of $\approx 10^{4}$--$ 3\cdot10^{7}$\,K, while for
temperatures lower than $10^{4}$\,K heating is the dominant
process. Nevertheless, these length scales are so large (in the order
of 10$^{4}$\,pc) that they exceed the TS distance by far, especially
in the tail. Thus they may be neglected to first order. If the thermal
pressure starts to dominate (beyond $T\approx 3\cdot10^{7}$\,K),
heating is by far the dominant process, with characteristic scales
of around 1\,pc and linearly (Eq.~(\ref{eq:c5})) increasing with
temperature.

Because of the huge ram pressure of the stellar wind, we
can from a dynamical point of view safely neglect heating and
cooling inside the TS, however, as it does not influence the polytropic
expansion of the stellar wind. Therefore, we can neglect the thermal
pressure compared with the ram pressure at the TS.

From the discussion above, we conclude that heating and cooling are
dynamically important only in the shocked and unshocked ISM, while for
the stellar wind it can be neglected, except in the extended tail
regions.

\subsection{Temperature and pressure}

The same considerations hold for the shocked temperature and pressure,
as discussed in Sect.~\ref{sec:p1}. We only have to assume that the
shocks are not radiative, meaning\ that the $\mathcal{P}_{i}=0,$ as
discussed in Sect.~\ref{sec:rdh}, which means that the thickness of
the shocks is so small that no collisions occur and the
excitation of atoms is negligible as well.  This is the case for astrospheres with
stationary
shocks because the material has enough time and
space to flow around the obstacles. For colliding shocks or shocks in
binary systems, this is not true \citep{Reitberger-etal-2014b,Parkin-etal-2011}.

For the \sfl case with cooling or heating, the supersonic ISM flow hits
the BS in the same way as without cooling. Thus the temperature
jump at the BS is given by the ISM speed (Eq.~(\ref{eq:t1})) 
and also the \SA $\vartheta$, while for the pressure
(Eq.~(\ref{eq:p1})) the ISM density is additionally needed.

To summarize, measuring the temperature along the shock as well gives the
interstellar wind speed when cooling and heating are active.

\section{Numerical models}

\begin{table}[t!]
  \centering
  \begin{tabular}{lllrr}
    Type & parameter     & unit       & value \\
\hline
    SW   & temperature   & K          & 453   \\
    SW   & speed         & km/s       & 2500  \\
    SW   & number density& cm$^{-3}$  & 24     \\
    ISM  & temperature   & K          & 10\,000    \\
    ISM  & speed         & km/s       & 80  \\
    ISM  & number density& cm$^{-3}$  & 11     \\
\hline
  \end{tabular}
  \caption{Initial values for the stellar wind (SW) at the inner
    boundary $0.05$\,pc and the
    interstellar medium (ISM) at ``$\infty$''. }
  \label{tab:c1}
\end{table}

In Fig.~\ref{fig:n-ph} we show the number density in the ``ecliptic''
for both the pH- and CH-model. A stationary equilibrium for the
pH-model is reached after roughly 1.5\,Myr, while it takes about
after 2.5\,Myr for the CH-model to reach stationary state. We
compare the two models after 1.5\,Myr to emphasize the differences in
time evolution.

\begin{figure*}[t!]
%  \centering
  \includegraphics[width=0.415\textwidth]{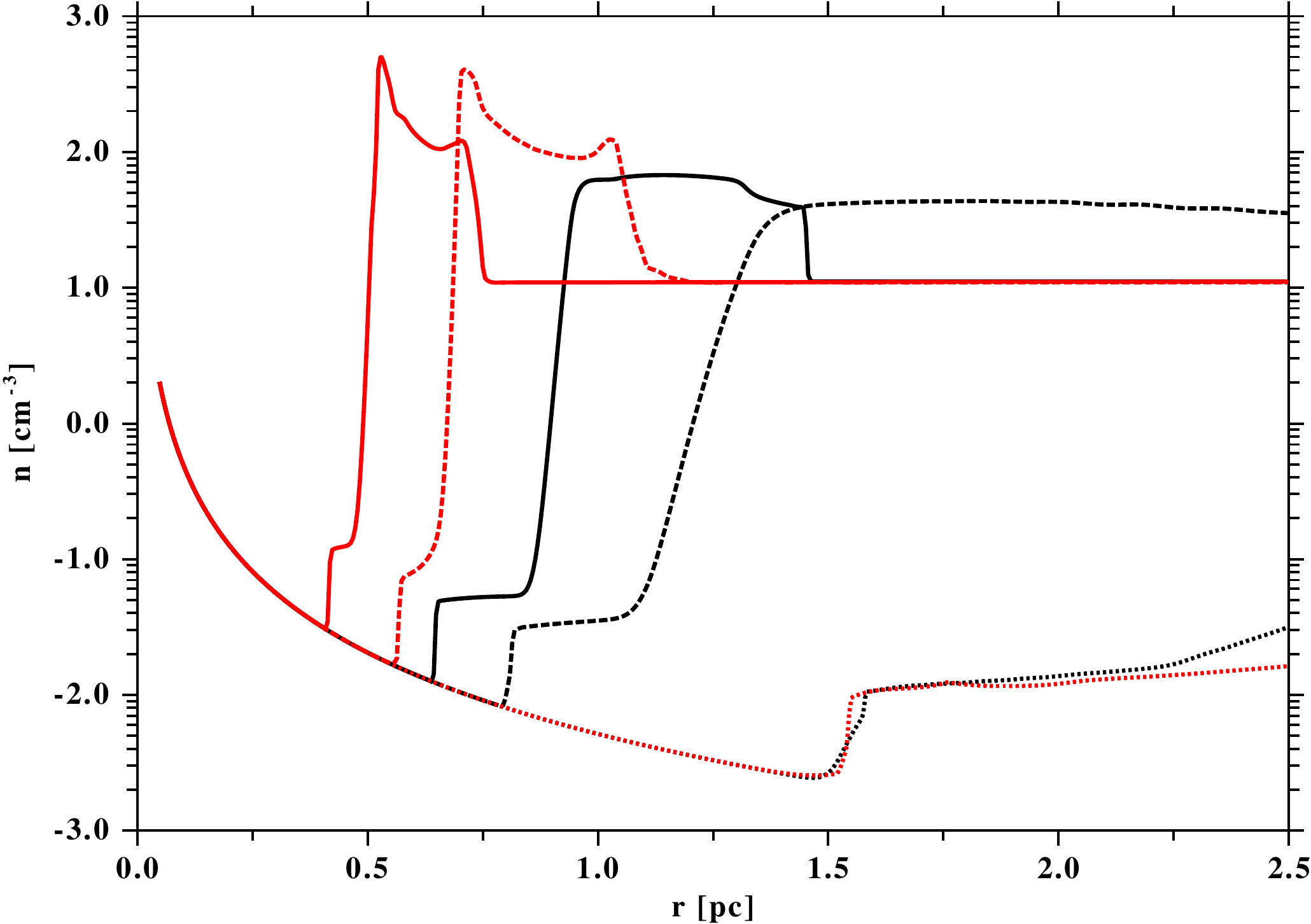}\hfill
  \includegraphics[width=0.415\textwidth]{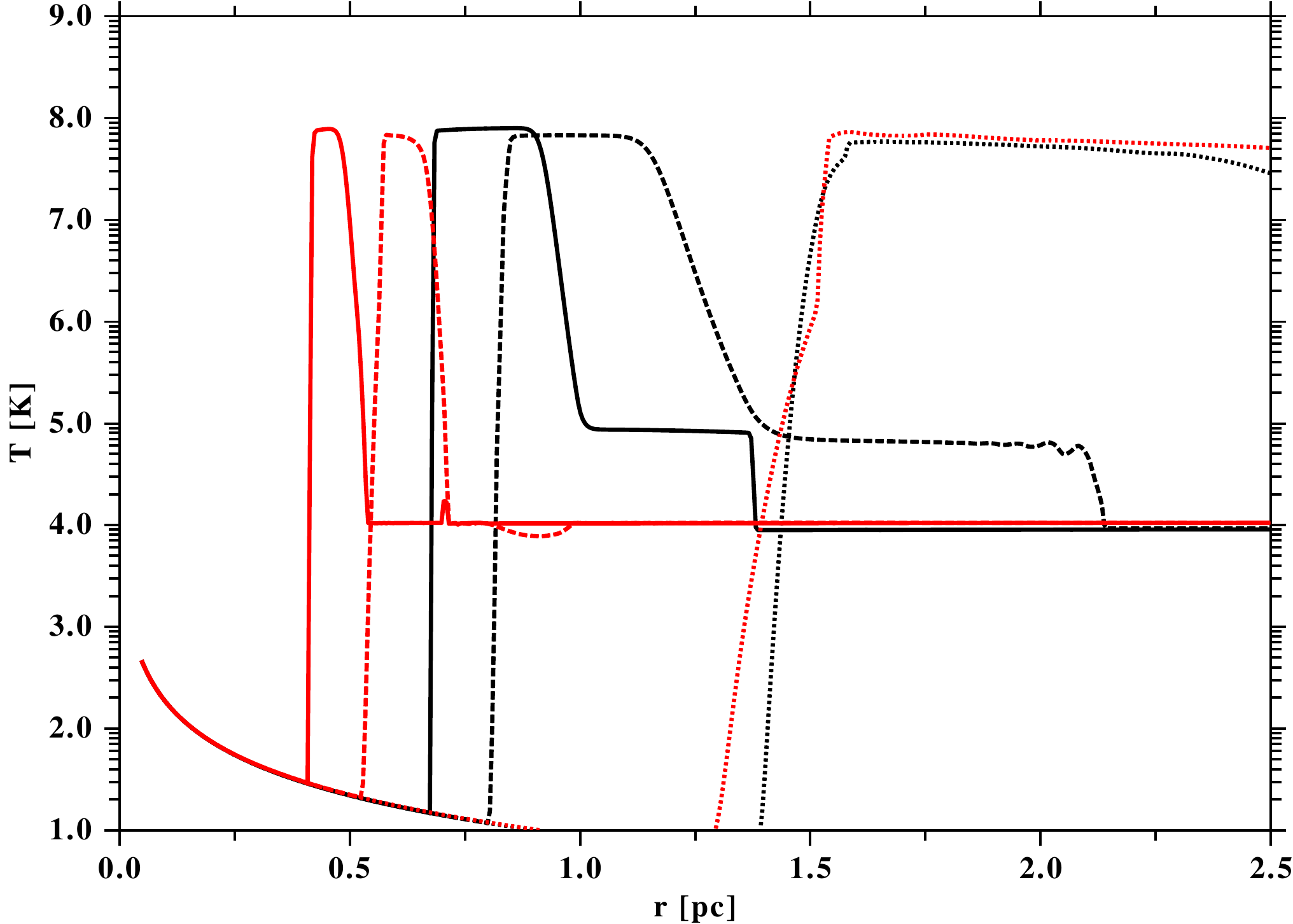}\\
  \includegraphics[width=0.415\textwidth]{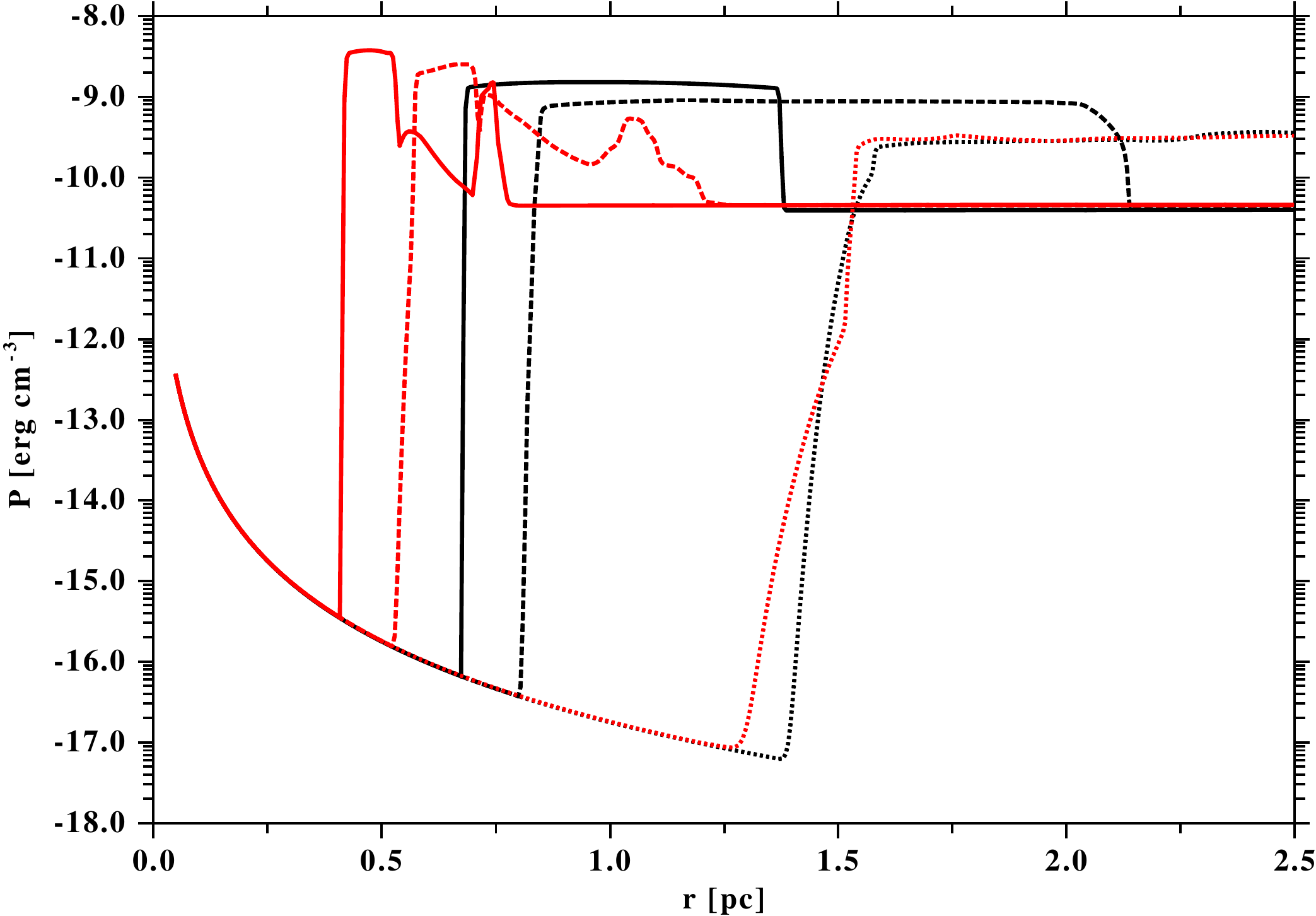}\hfill
  \includegraphics[width=0.415\textwidth]{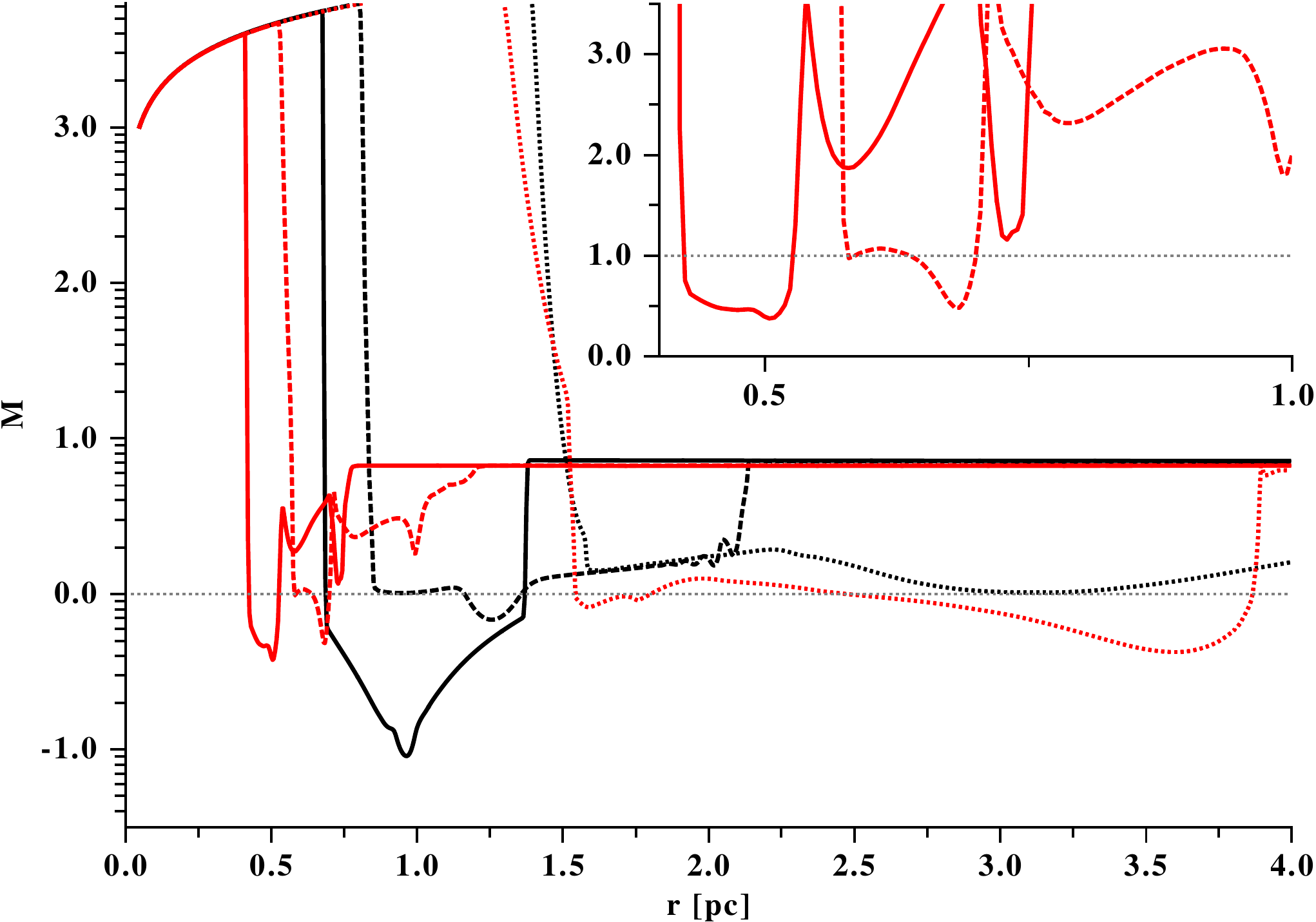}
  \caption{Number density, temperature, thermal pressure, and the
    Mach number. The black lines
    describe them for the model without and the red lines the
parameters for
    the model with cooling and heating. The solid lines show the parameters
    along the stagnation line (i.e.\ 0$^{\deg}$), while the dashed lines
    are those along the pole (90$^{\deg}$), and the dotted lines along the
    tail direction (180$^{\deg}$). The upper left panel shows
    that the number density for the model without cooling and heating
    jumps at the TS and BS by a factor of four, while for the CH model
    the density at the TS also increases by a factor of four, but at
    the BS it increases by more than a factor of 10. In both cases the
    temperature jumps by orders of magnitudes at the shocks (upper
    right panel). The lower left panel shows the thermal pressure,
    which is discontinuous at the shocks, but in both cases more or
    less continuous at the AP along the stagnation line. The lower
    right panel shows the Mach number.  The inlet is a linear plot in
    the Mach number between 0 and 3. It shows only the Mach numbers in
    the CH model to present its structure.  The CH model is presented
    after 150\,kyr, while the pH model is fully developed (kyr).  For
    more details see text.}
\label{fig:ab}
\end{figure*}

In Fig.~\ref{fig:ab} some important quantities along a radius vector
in the ecliptic plane ($\vartheta=0^{o}$) are shown. The pH and CH
model are asymmetric, therefore lines in the figure are sufficient to
show the essential variations. While in the pH model the density at
the TS jumps by a factor four (strong shocks), it remains constant up
to the AP. There it increases by orders of magnitudes to the shocked
interstellar value, and finally decreases by a factor $\approx3.8$ at
the BS to the interstellar value. The ISM number density at the AP is
slightly higher at the AP because here the velocity tends to zero and
the gas is piled up.

For the CH model the behavior between the TS and the AP is
analogous. The ISM gas density is lower than in the pH model because the cooling also takes place in the ISM, and a new equilibrium state of the ISM is attained just inside the outer
boundary conditions. 

The CH model again shows a pile up of the ISM gas at the AP,
while it almost decreases to the new ISM value shortly after the
AP. Then the ISM number density increases further by a factor 50. This
compression is about 12 times stronger than the compression for a pure
gas dynamic shock, where we have defined the compression ratio $s$ as
the ratio of the number density in front of the shock and behind
it. In the CH model the increase of the density to its highest value
is reached after 0.05 pc, and therefore we took the value at this position
to calculate the compression ratio. This behavior is also described in
\citet{Mackey-etal-2015b}, and it leads to a division of the outer
astrosheath into a cool and hot part (see below).

The distance between the first peak as seen from the BS to the AP is
$\approx 0.05$\,pc in the CH model, which is roughly the cooling
length. The distance between the AP and the BS is roughly 0.2\,pc in
the CH model and 0.4\,pc in the pH model. The distance between the TS
and the AP in both models is approximately 0.2\,pc. Together, the
distance between the TS and BS is approximately 40\% that of the
stellar-centric distance to the TS. This again shows that the thin-shell approximation is not valid because the outer shell (between TS
and BS) has almost the same size as the inner shell (between origin
and TS). In addition, the density increases by a factor of more
than
four at the flanks at 90$^{\deg}$ and in the tail
direction 180$^{\deg}$  , and the thickness of the outer astrosheath is almost the same in
both models.

 As shown in the lower left panel Fig.~\ref{fig:ab}, the thermal
 pressure is almost the same in the inner and outer astrosheath and
 is continuous at the AP, as required for a tangential
 discontinuity. This holds also true for the other two directions.

 The temperature is displayed in the upper right panel of Fig.~\ref{fig:ab}. It increases by orders of magnitudes at the TS in both
 models and remains constant in the pH model until it reaches the AP,
 where it decreases by orders of magnitudes to the shocked ISM value,
 and finally decreases at the BS to unshocked values. The sudden increases at
 the AP are slightly smeared out for the TS at the polar direction,
 while the BS does not exist in the tail direction. The increase of
 the temperatures at the TS in the CH model is almost the same as for
 the pH model, but then it decreases strongly at the AP to the
 unshocked ISM value, shows a peak in the direction of the BS, and finally
 returns to the unshocked value.  In the polar direction, the
 increase toward the AP in the inner astrosheath is similar to that
 for the pH model, but after the AP it directly drops to the unshocked
 value in the ISM. There is no longer a BS at
this position. In the
 tail direction, the temperature remains at the high value after the
 shock for both models.

The Mach numbers are shown in the lower right panel of Fig.~\ref{fig:ab}. Those of the CH model in the nose direction and
 90$^{\deg}$ are separately displayed in the inset. Here the scale on the vertical
 axis is linear to show the complexity. The Mach number along the
 stagnation line drops after the TS to values below one and decreases
 further to the astropause. There it quickly increases and shows some
 structure in the outer astrosheath, namely a hot outer astrosheath
 (HOA) and a cool outer astrosphere (COA). These structures differ in
 the HOA and COA. Moreover, it depends on the choice of the CHF. Thus
 a detailed comparison between the CHF is required, which may also
 be reflected in the observations, leading to some information
 about the composition of the ISM because the CHF depends on it
 \citep[e.g.,][]{Sutherland-Dopita-1993}.

\begin{figure*}[t!]
  \centering
  \includegraphics[width=0.45\textwidth]{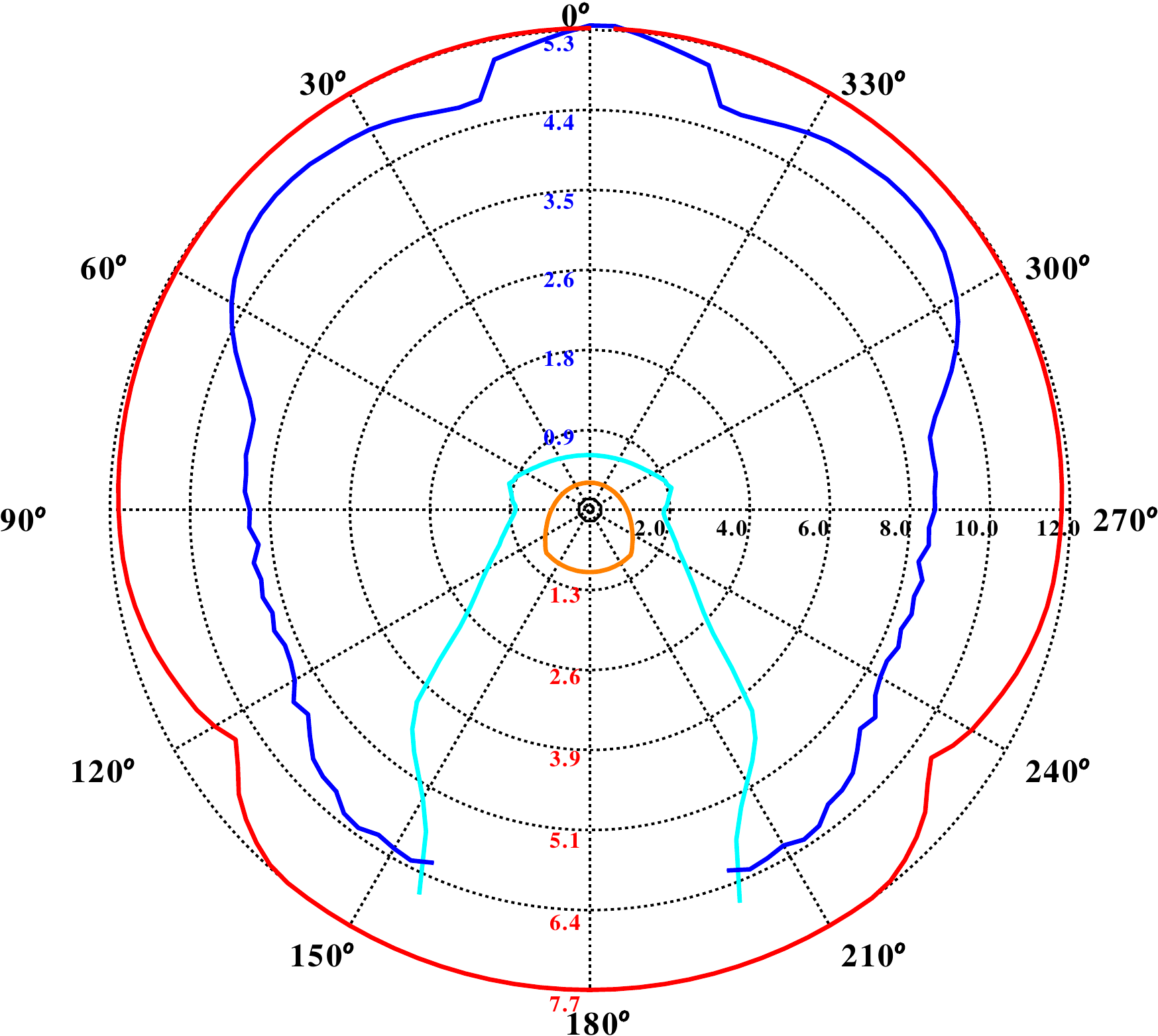}\hfill
  \includegraphics[width=0.45\textwidth]{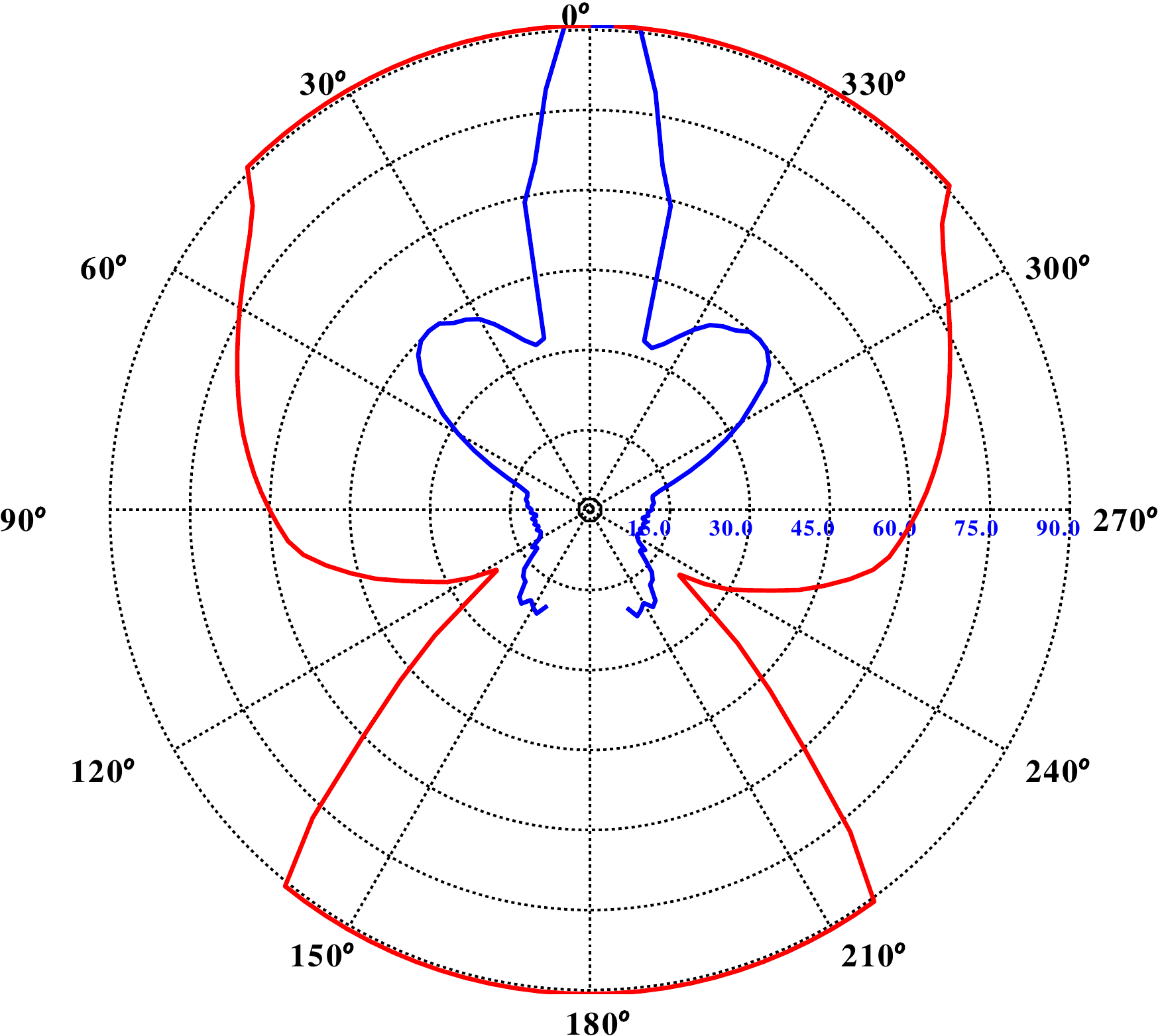}\\
  \includegraphics[width=0.45\textwidth]{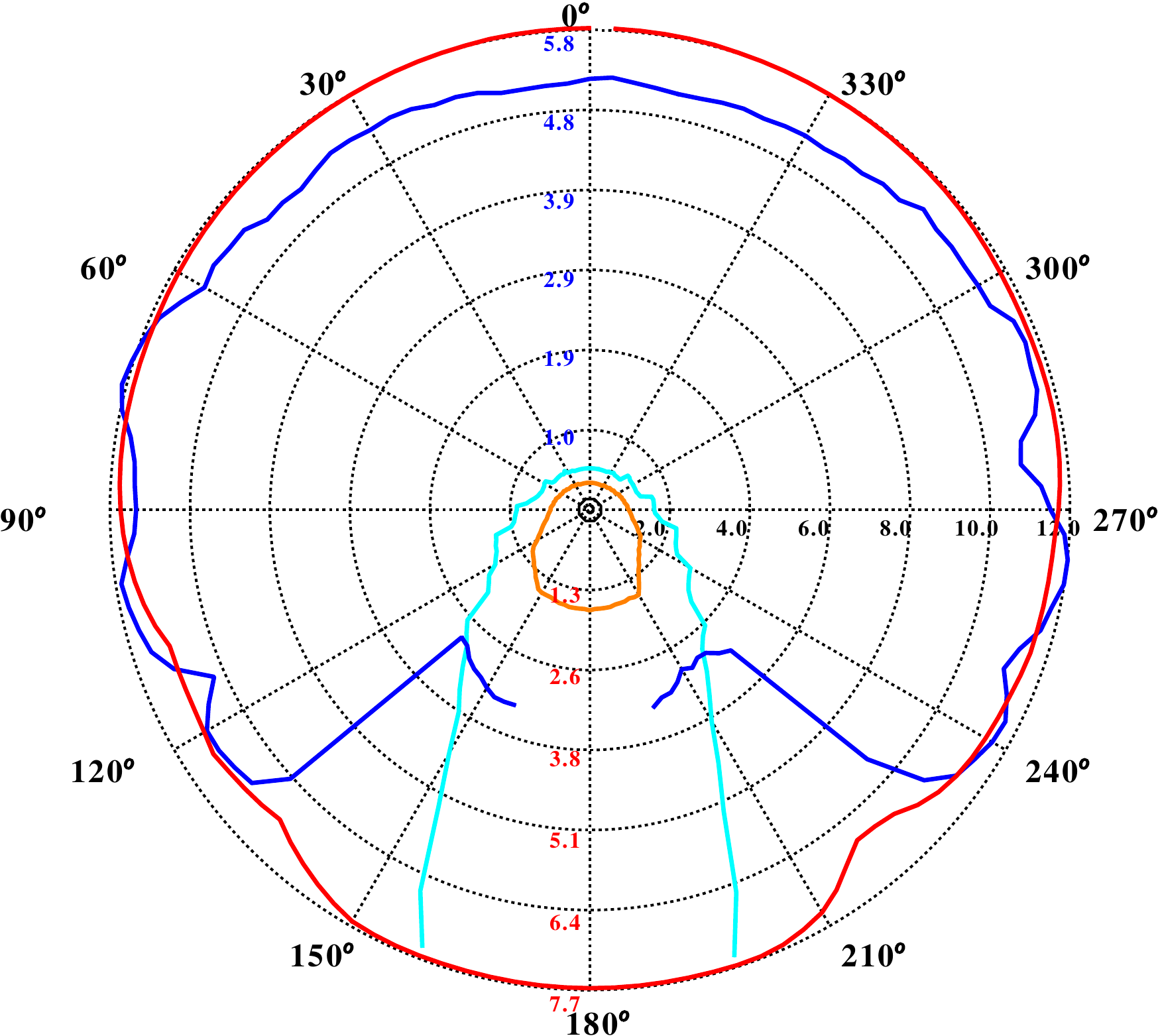}\hfill
  \includegraphics[width=0.45\textwidth]{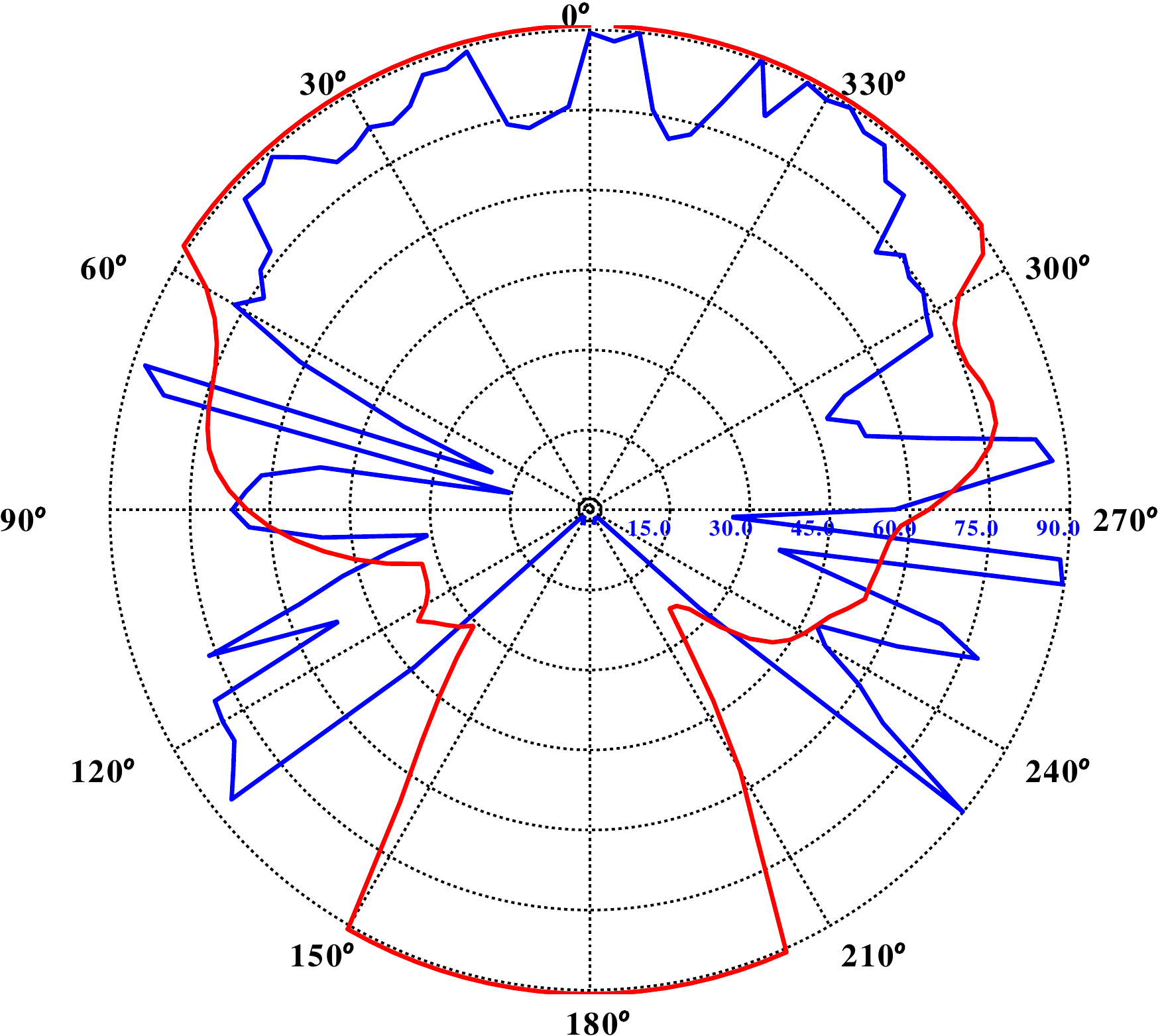}
  \caption{Left panels: the shock distance depending on the
    stellar-centric angle $\Phi$ counted from the inflow direction. Cyan shows the distance of the BS, orange that of the TS. Blue shows the variation of the
temperature along the BS and red that of the TS. Right panels: the \SAs at the BS (blue) and the TS (red). The upper two panels display the quantities for the pH-model, the lower panels those for the CH-model. For more details see
text.}
\label{fig:rund}
\end{figure*}

From this discussion we can extract much information about
the shock structure: In Table~\ref{tab:c1} we list the inner boundary values
for the stellar wind of $\lambda$ Cephei at 0.03\,pc with those at the
outer boundary at 12\,pc$\hat{=} \infty$ for the ISM. The polytropic index is fixed in the entire integration
region to $\gamma = 5/3$. Using these set of parameters, we can
determine the sound speeds, Mach numbers, and pressures as well as the
distances to the TS in nose and tail direction. The latter two are
compared with the modeled distances. The results are shown in
Table~\ref{tab:c3} together with the compression ratios
($s = \rho_{2} /\rho_{1}$). Finally, we estimated the values of the
stellar wind at the TS, using the $r$ dependence for polytropic
expanding stellar wind (i.e.,  $\rho \propto r^{-2}$,
$T\propto r^{-4/3}$ , $P\propto r^{-10/3}$, and $M\propto r^{2/3}$),
which are displayed in Table~\ref{tab:c3}, and those for the
interstellar parameters at the BS, shown in Table~\ref{tab:c4}. Finally we display
in Table~\ref{tab:c5} the \SAs and \FAs at the intersection of the
sonic lines with the TS and BS.

In Tables~\ref{tab:c4} and~\ref{tab:c5} we show the results from the pH
and CH models. They agree well with the
analytic results. The tables also show that the values at the TS
and BS do not change much from one model to the next. In the outer
astrosheath, the flow behaves different in both models. This
means that the analytic values shown in Tables~\ref{tab:c1} to~\ref{tab:c5}
can be used as a first estimate for an astrosphere.
\begin{table}[t!]
  \centering
  \begin{tabular}{lllrr}
        Distance  & analytic     & pH      & CH  \\
\hline
$R_{TS,n}$  & 0.67 & 0.65 & 0.65\\
$R_{AP}$    & -    & 0.90 & 0.80\\
$R_{BS}$    & -    & 1.45 & 1.16\\
$R_{TS,t}$  & 2.81 & 1.58 & 1.56\\
  \end{tabular}
  \caption{Characteristic distances. All values are given in
    units of pc. The TS distance in the nose
    direction remains the same  in the analytic approach and in the pH
    and CH models. The AP distances 
    for the CH model are smaller because the thermal pressure in the outer
    astrosheath is decreased by the cooling. This holds also true
    for the BS distance. The TS distance in the tail direction of the
    analytic calculation deviates by roughly a factor 2 compared to
    both numerical models, which gave almost the same value.
  }
  \label{tab:c2}
\end{table}

\begin{table*}[t!]
  \centering
  \begin{tabular}{ll|r@{.}l|r@{.}lr@{.}lr@{.}l}
 &&&&\multicolumn{2}{c}{analytic} & \multicolumn{2}{c}{pH model}
 &\multicolumn{2}{c}{CH model} \\
  parameter     & unit      &  \multicolumn{2}{c|}{$x= 1$}   & \multicolumn{2}{c}{$x=2$} &  \multicolumn{2}{c}{$x= 2$}   & \multicolumn{2}{c}{$x=2$}     \\
\hline
$M_{x,ism}$    &            & 9&28    & 0&46 &  0&51 & 0&81\\
$n_{x,ism}$    & cm$^{-3}$  & 11&00   &45&52 & 43&23 & 40&26\\
$v_{x,ism}$    & km/s       & 80&00   &20&70 & 20&36 & 21&86\\
$T_{x,ism}$    & K          & 9&00$\cdot10^{3}$ & 2&5$\cdot10^{5}$ & 8&11$\cdot10^{4}$&
1&5$\cdot10^{5}$ \\
$P^{therm}_{x,ism}$ & dyne  &4&00$\cdot10^{-11}$ &2&50$\cdot10^{-9}$ &1&25$\cdot10^{-9}$
& 1&23$\cdot10^{-9}$\\
$P^{ram}_{x,ism}$   & dyne  &1&18$\cdot10^{-9}$ & 3&05$\cdot10^{-10}$
& 2&87$\cdot10^{-10}$& 3&08$\cdot10^{-10}$\\
  \end{tabular}
  \caption{ Parameters of the ISM in front of the
    shock (upstream $x=1$) and behind it (downstream $x=2$).
}
  \label{tab:c4}
\end{table*}

The CHF appears only in the energy equation, not in the momentum
equation. The latter is weakly influenced through thermal pressure
changes induced by the energy equation. The reason for the lack of
momentum changes is that the energy loss is a scalar that is radiated
in all directions, without changing its sign. The momentum is a vector
quantity that in the rest frame of a particle also radiates in all
directions, whith two movements cancelling each other. This means for
a photon emitted in one direction that there will be on average one
photon with the same moment emitted in the opposite direction. Thus
the momentum loss can be neglected.

The cooling and heating also introduce a notorious instability, as
already discussed in \citet{Schwarz-etal-1972} and
\citet{Scherer-etal-2015a}. This instability results in wiggles
along the BS. By comparison with observations, the wavelength of these
wiggles may be used to determine the CHFs in
mind, if it can be assume that the wiggles depend in a characteristic
way on the magnitude of the CHF.

Finally, we applied the hypersonic formula for the dependence between
the temperature and the inflow speed and \SA, Eq.~(\ref{eq:t1}), to
estimate the \SA. In Fig.~(\ref{fig:rund}) we display the shock
distances for both the TS and the BS along the stellar-centric angle
$\Phi$ counted from the inflow direction. The left panels of
Fig.~\ref{fig:rund} show the shock distances and the temperatures along them, but in the shocked
region, the TS and BS. The shocked temperatures after the TS in the inflow and in the tail
direction are almost equal, in agreement with the Rankine-Hugoniot
relations. In the right panels the \SAs along both shocks for both
models are shown. In the upper left panel of Fig.~\ref{fig:rund} the shocked temperature (blue line) upstream of the
BS at roughly 15$^{\deg}$ and 345$^{\deg}$ has a depression that
strongly decreases to values of around 30$^{\deg}$. When the \SA reaches 60$^{\deg}$ , the
decreases becomes even larger and the hypersonic approximation is no
longer valid. This occurs at the sonic points, beyond which a weak
shock solution applies.

The figure also shows a similar feature at the TS shortly after 40$^{\deg}$ (320$^{\deg}$), where the shocked Mach
number also becomes 1 and the \SA $\approx 63^{o}$. In the tail direction,
at about 150$^{\deg}$ (210$^{\deg}$), another depression is visible, after
which we find a strong increase of the \SA to almost 90$^{\deg}$. These
depletions occur at the triple point and the increase at the
Mach disk to the required 90$^{\deg}$ \SA at the downwind axis.

First the astrosheath is divided into a cool and hot part
\citep[see also][]{Mackey-etal-2015b}. The shocked Mach number is everywhere greater than one in both
outer astrosheaths. This is in
contrast to the pH model, where the Mach number
is lower than one in the nose direction. Thus the sonic line between the BS and AP
disappears, and the outer astrosheath is supersonic everywhere. In the
inner astrosheath, the sonic line starts at the same point as in the
pH case, but then bends in the middle of the inner astrosheath into
the tail direction, where it then ends in the AP. The triple point is
still existent, but the bending of the shock is much smoother. The
reflected shocks vanish. Because the triple points still exists, the
Mach disk remains.

While in the HOA the number density jump after the shock is similar to
that in the pH case, it decrease toward the COA, to increase to much
higher values in the COA than for the pH model. In front of
the AP a proton wall is created in this way because the normal velocity decreases
to zero at the AP (no mass transport through it).

Similar for the temperature: it increases after the shock passage in the
HOA and then quickly becomes lower when the cooling starts to
operate. It almost reaches the same values as those in the ISM. The
pressure in the HOA and COA remains more or less constant, as can be
inferred from the ideal gas equation $P=2nkT$. Accordingly, when the density
increases, the temperature has to decrase to keep the
pressure constant. This behavior is shown in Fig.~\ref{fig:ab}. 

\begin{table*}[t!]
  \centering
  \begin{tabular}{ll|r@{.}lr@{.}l|r@{.}lr@{.}l|r@{.}lr@{.}l}
& &\multicolumn{4}{c}{analytic} & \multicolumn{4}{c}{pH model}
 &\multicolumn{4}{c}{CH model} \\
  parameter     & unit      &  \multicolumn{2}{c}{$x= 1$}   & \multicolumn{2}{c}{$x=2$} &  \multicolumn{2}{c}{$x= 1$}   & \multicolumn{2}{c}{$x=2$} &  \multicolumn{2}{c}{$x= 1$}   & \multicolumn{2}{c}{$x=2$}    \\
\hline
    $cw_{x,sw,n}$ & km/s     & 0&45    & 1398& &0&45 & 1005& & 0&45 & 1029&\\
    $cw_{x,sw,t}$ & km/s     & 0&17    & 1398& &0&31 & 108&71 & 0&22 & 151&\\
    $M_{x,sw,n}$  &           & 5602&    & 0&45 &5595& &0&62 & 5550& &    0&62\\ 
    $M_{x,sw,t}$  &           & 15049&   & 0&45 &9912& & 9&41 & 11160& & 8&51  \\ 
    $n_{x,sw,n}$  & cm$^{-3}$ & 1&13$\cdot10^{-2}$ &
    4&15$\cdot10^{-2}$ & 1&41$\cdot10^{-2}$ &4&8$\cdot10^{-2}$
    & 2&99   $\cdot10^{-2}$ &  1&87   $\cdot10^{-1}$\\
    $n_{x,sw,t}$  & cm$^{-3}$ & 5&81$\cdot10^{-4}$&2&32$\cdot10^{-3}$
    & 7&11$\cdot10^{-3}$ & 1&05$\cdot10^{-2}$ &3&40$\cdot10^{-3}$ &1&02$\cdot10^{-2}$\\
    $v_{x,sw,n}$  & km/s      & 2500&    & 625&   &2500& & 623& &2500& &886&  \\
    $v_{x,sw,t}$  & km/s      & 2500&    & 625&   &2500& & 1023&
    &2500& &909&  \\
    $T_{x,sw,n}$  & K         & 1&45$\cdot10^{1}$  & 1&42$\cdot10^{8}$
    & 1&45$\cdot10^{1}$  & 7&8$\cdot10^{7}$& 1&56$\cdot10^{1}$  & 7&5$\cdot10^{7}$\\ 
    $T_{x,sw,t}$  & K         & 2&01$\cdot10^{0}$  & 1&42$\cdot10^{8}$
    & 9&01$\cdot10^{0}$  & 5&7$\cdot10^{7}$ & 8&93$\cdot10^{0}$  & 7&22$\cdot10^{7}$\\ 
    $P^{therm}_{x,sw,n}$  & dyne      & 6&75$\cdot10^{-17}$&
    2&65$\cdot10^{-9}$ & 6&47$\cdot10^{-17}$ & 1&27$\cdot10^{-9}$ &1&89$\cdot10^{-17}$&2&01$\cdot10^{-9}$\\ 
    $P^{therm}_{x,sw,t}$  & dyne      & 4&83$\cdot10^{-19}$&
    1&37$\cdot10^{-10}$ & 9&46$\cdot10^{-18}$ &1&20$\cdot10^{-10}$& 1&60$\cdot10^{-17}$& 1&44$\cdot10^{-10}$\\ 
    $P^{ram}_{x,sw,n}$    & dyne      &
    1&18$\cdot10^{-9}$&2&94$\cdot10^{-10}$
    &1&41$\cdot10^{-9}$&2&98$\cdot10^{-10}$ 
    &2&99$\cdot10^{-9}$ & 2&35$\cdot10^{-11}$\\
    $P^{ram}_{x,sw,t}$    & dyne      &
    5&81$\cdot10^{-11}$&1&45$\cdot10^{-11}$ 
    & 7&011$\cdot10^{-10}$&6&52$\cdot10^{-12}$
    & 3&40$\cdot10^{-10}$&1&35$\cdot10^{-11}$\\
  \end{tabular}
  \caption{Stellar wind parameters in front of the shock (index 1)
    and behind it (index 2). The indices $n\text{ and }t$ denote the nose and tail
  direction.}
  \label{tab:c3}
\end{table*}

For the CH-model the \SA of the TS behaves similarly, but that of the BS
is more or less arbitrary. The reason for the latter is that the HOA
is very thin around the inflow direction, and determining the
temperature is difficult. The flow becomes even hotter in the flanks
of the astrosphere, which is caused by the interaction of flow from
the nose direction and the directly shocked stellar wind in the outer
astrosheath regions.

The curves shown in Fig.~\ref{fig:rund} are ragged because of the
low resolution in the $\Phi$ direction and the strong dependence on
$\arcsin$.

\begin{table}[t!]
  \centering
  \begin{tabular}{ll|rrr}
 &&\multicolumn{1}{c}{analytic} & \multicolumn{1}{c}{pH model}
 &\multicolumn{1}{c}{CH model} \\
  parameter          & unit      & value\\
\hline
$\vartheta_{sl,ism}$ & degree    & 63.43 & - & - \\
$\vartheta_{sl,sw}$  & degree    & 63.11 & - & - \\
$\varphi_{sl,ism}$   & degree    & 36.87 & - & - \\
$\varphi_{sl,sw}$    & degree    & 35.88 & - & - \\
$s_{sw,n}$           &           & 4.00  &  3.93 & 3.66 \\
$s_{sw,t}$           &           & 4.00  & 3.85  & 3.57 \\
$s_{ism}$            &           & 3.87  & 3.63 & 2.99  \\

  \end{tabular}
  \caption{Derived shock parameters: The \SA $\vartheta_{sl,i}$ and
\FA $\varphi_{sl,i}$ at the intersection of the sonic line and the
TS ($i = sw$) and that at the BS ($i = ism$) are displayed together with
the compression ratios $s_{j}$ at the TS in nose direction ($ j = n$), the tail
direction $j = t$ and at the BS $j = ism$. The sonic lines are poorly
defined in the CH model, and therefore  neither the \SA nor the \FA  are
given here. In the pH-model they are difficult to determine because the \SA\ smoothly evolves from 90$^{\deg}$ to the lowest
values at the triple point. The easiest way therefore is to calculate them
analytically. 
}
  \label{tab:c5}
\end{table}
\begin{table}[t!]
  \centering
  \begin{tabular}{lllrr}
        Type & parameter     & unit       & value \\
\hline
        SW  & sound speed    & km/s       & 2.75   \\
        ISM & sound speed    & km/s       & 8.62   \\
        SW  & Mach number    &            & 1001.46 \\
       ISM  & Mach number    &            & 9.28   \\
        SW  & pressure       & dyne       & 3.70$\cdot10^{-13}$ \\
       ISM  & pressure       & dyne       & 4.56$\cdot10^{-11}$ \\
  \end{tabular}
  \caption{ Derived parameters for the stellar wind (SW) at the inner (at
    0.05\,pc) and for the interstellar medium (ISM) at the outer
    boundaries (at ``$\infty$'').} 
  \label{tab:c6}
\end{table}
 
\section{ \Lsc shock structure including cooling and heating}

\begin{figure*}[t!]
  \centering
    \includegraphics[width=0.9\textwidth]{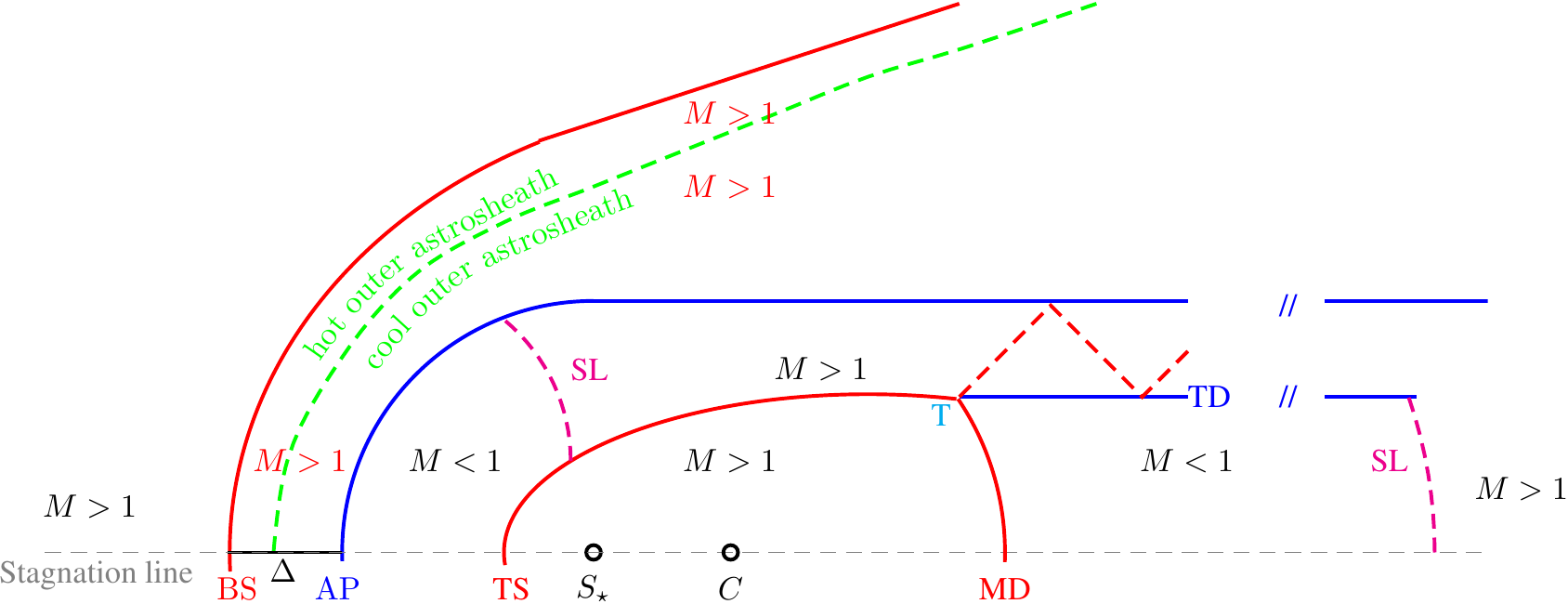}
    \caption{Sketch of an astrosphere with strong cooling and heating
      effects. The sonic lines are different from those in
      Fig.~\ref{rh:1}, and in some region the Mach number has changed.
      The boundary between the hot and cool outer astrosheath is
      marked with a dashed green line. The changes in the Mach number
      are indicated in red.}
\label{fig:ac}
\end{figure*}

We combine all our findings in the sketch shown in
Fig.~\ref{fig:ac}.

The structure of astrospheres including heating and cooling differs from
those without it because the structure of the former strongly depends on the
magnitude of these effects. They may not play a role at all, as is shown by the example of the heliosphere, where the cooling and heating
lengths are much longer than the dimension of the
heliosphere. However, for massive stars with huge astrospheres, the
characteristic cooling and heating lengths are shorter than the
characteristic scales, like the distances between the BS and AP.

\section{Conclusions}

Astrospheres with cooling and heating in general have a different
structure from those without it. In contrast to \citet{Mackey-etal-2015},
we used the Mach number as the dynamically most important parameter to
describe the structure of such an astrosphere. We found that the
outer astrosheath is separated into a cool and a hot part, but there is
more fine structure when using the Mach number as structuring
parameter. The latter is interesting for acceleration processes of
energetic particles \citep{Scherer-etal-2015a}.

We gave an \ide analysis of a pure hydrodynamical \sfl
(pH-) model and compared it with a model that also included heating and
cooling effects (CH-model). We compared some characteristic parameters
of the two models with those obtained by the Rankine-Hugoniot relations
for a single-species fluid. We found good agreement with the
pH model, while for the CH model only the relations along the TS give
reliable results.

We also showed that the thin-shell approximation is not valid in
this context because the inner and outer astrosheath are
comparable everywhere with the region inside the TS. Moreover, from theoretical
considerations we are led to the same result because the \SA is not
90$^{\deg}$ and no statement about the \FA can be given in general.

We discussed the application of the Rankine-Hugoniot relations to the
models with and without cooling and heating. With these relations the
conditions at the TS and BS can be estimated. We also showed that in
the pH model only the TS distance can be estimated with good accuracy,
while the AP and BS distances are difficult to determine because they
do not have an explicit $r$-dependence. For the CH-model, we can
estimate at least the thickness of the astrosheath using the cooling
lengths defined above.

We also showed that for hypersonic ISM flows the temperature along the
BS depends only on the unshocked ISM speed and the \SA
$\vartheta$. Thus, measuring a temperature variation along the BS can
give the unshocked ISM speed at the stagnation line and the \SAs
along the BS.  Because of the high Mach number (hypersonic) stellar
wind flow, the thermal pressure (or equivalently temperature) does not
play a dynamical role until the flow reaches the TS. Directly after
this, the thermal pressure (or temperature) is determined by the flow
speed and the \SAs along the TS. For modeling purposes the
knowledge of the stellar wind speed and densities at a given position
are therefore sufficient to model the astrosphere. Nevertheless, an approximate
value has to be allocated in most of numerical codes, which should be
low enough.  We also showed that the compression ratio at the BS
in the CH model is much higher than in the pH model. The velocity
divergence at the TS is similar, but is much
more compressed in the CH model.

We also demonstrated that the cooling or heating is mainly
effective in the outer astrosheath because there the densities are
high and speeds low enough to enable an efficient cooling, while in
the inner astrosheath the densities are very low and the bulk
speed is very high, so that the cooling length scales are much longer
than the TS distance, and thus the cooling and heating do not
operate efficiently.

\begin{acknowledgements}
  KS, HF, JK, and TW are grateful to the
  \emph{Deut\-sche For\-schungs\-ge\-mein\-schaft, DFG,\/} funding the
  projects FI706/15-1 and SCHE334/10-1. DB and KW were supported
  by the DFG Research Unit FOR 1254.  KS and HF  appreciate
  discussions at the team meeting ``Heliosheath Processes and
  Structure of the Heliopause: Modeling Energetic Particles, Cosmic
  Rays, and Magnetic Fields'' supported by the International Space
  Science Institute in Bern, Switzerland.
\end{acknowledgements}
%\bibliographystyle{aa}
%\bibliography{\HOME/Tex/references}

\begin{thebibliography}{97}
\expandafter\ifx\csname natexlab\endcsname\relax\def\natexlab#1{#1}\fi

\bibitem[{{Alexashov} {et~al.}(2004){Alexashov}, {Chalov}, {Myasnikov},
  {Izmodenov}, \& {Kallenbach}}]{Alexashov-etal-2004}
{Alexashov}, D.~B., {Chalov}, S.~V., {Myasnikov}, A.~V., {Izmodenov}, V.~V., \&
  {Kallenbach}, R. 2004, \aap, 420, 729

\bibitem[{{Alouani-Bibi} {et~al.}(2011){Alouani-Bibi}, {Opher}, {Alexashov},
  {Izmodenov}, \& {Toth}}]{Alouani-Bibi-etal-2011}
{Alouani-Bibi}, F., {Opher}, M., {Alexashov}, D., {Izmodenov}, V., \& {Toth},
  G. 2011, \apj, 734, 45

\bibitem[{{Arthur}(2007)}]{Arthur-2007}
{Arthur}, S. 2007, Rev. Mex. Astron. Astrophys., 30, 64

\bibitem[{{Arthur}(2012)}]{Arthur-2012}
{Arthur}, S.~J. 2012, \mnras, 421, 1283

\bibitem[{{Baranov} {et~al.}(1971){Baranov}, {Krasnobaev}, \&
  {Kulikovskii}}]{Baranov-etal-1971}
{Baranov}, V.~B., {Krasnobaev}, K.~V., \& {Kulikovskii}, A.~G. 1971, Soviet
  Physics Doklady, 15, 791

\bibitem[{Barnette(1993)}]{Barnette-1993}
Barnette, D. 1993, Program SHOCKS: Quickly Estimating Super- and Hypersonic
  Inviscid Flow Parameters., Tech. rep., Sandia National Laboratories,
  Albuquerque, New Mexico

\bibitem[{{Ben-Dor}(2007)}]{Ben-Dor-2007}
{Ben-Dor}, G. 2007, {Shock Wave Reflection Phenomena} (Springer
  Science+Business Media)

\bibitem[{Blasi(2013)}]{Blasi-2013}
Blasi, P. 2013, Astron. Astrophy. Rev., 21, 1

\bibitem[{{Bouquet} {et~al.}(2000){Bouquet}, {Romain}, \&
  {Chieze}}]{Bouquet-etal-2000}
{Bouquet}, S., {Romain}, T., \& {Chieze}, J.~P. 2000, \apjs, 127, 245

\bibitem[{{Bucciantini}(2002)}]{Bucciantini-2002}
{Bucciantini}, N. 2002, \aap, 387, 1066

\bibitem[{{Bucciantini}(2014)}]{Bucciantini-2014}
{Bucciantini}, N. 2014, Astronomische Nachrichten, 335, 234

\bibitem[{{Bzowski} {et~al.}(2015){Bzowski}, {Swaczyna}, {Kubiak},
  {Sok{\'o}{\l}}, {Fuselier}, {Galli}, {Heirtzler}, {Kucharek}, {Leonard},
  {McComas}, {M{\"o}bius}, {Schwadron}, \& {Wurz}}]{Bzowski-etal-2015}
{Bzowski}, M., {Swaczyna}, P., {Kubiak}, M.~A., {et~al.} 2015, \apjs, 220, 28

\bibitem[{{Chashei} \& {Fahr}(2013)}]{Chashei-Fahr-2013}
{Chashei}, I.~V. \& {Fahr}, H.~J. 2013, Annales Geophysicae, 31, 1205

\bibitem[{{Chashei} \& {Fahr}(2014)}]{Chashei-Fahr-2014}
{Chashei}, I.~V. \& {Fahr}, H.~J. 2014, \solphys, 289, 1359

\bibitem[{{Courant} \& {Friedrichs}(1948)}]{Courant-Friedrichs-1948}
{Courant}, R. \& {Friedrichs}, K.~O. 1948, {Supersonic flow and shock waves}
  (New York: Interscience)

\bibitem[{{Cox} {et~al.}(2012){Cox}, {Kerschbaum}, {van Marle}, {Decin},
  {Ladjal}, {Mayer}, {Groenewegen}, {van Eck}, {Royer}, {Ottensamer}, {Ueta},
  {Jorissen}, {Mecina}, {Meliani}, {Luntzer}, {Blommaert}, {Posch},
  {Vandenbussche}, \& {Waelkens}}]{Cox-etal-2012}
{Cox}, N.~L.~J., {Kerschbaum}, F., {van Marle}, A.~J., {et~al.} 2012, \aap,
  543, C1

\bibitem[{{Dalgarno} \& {McCray}(1972)}]{Dalgarno-McCray-1972}
{Dalgarno}, A. \& {McCray}, R.~A. 1972, \araa, 10, 375

\bibitem[{{Decin} {et~al.}(2012){Decin}, {Cox}, {Royer}, {Van Marle},
  {Vandenbussche}, {Ladjal}, {Kerschbaum}, {Ottensamer}, {Barlow}, {Blommaert},
  {Gomez}, {Groenewegen}, {Lim}, {Swinyard}, {Waelkens}, \&
  {Tielens}}]{Decin-etal-2012}
{Decin}, L., {Cox}, N.~L.~J., {Royer}, P., {et~al.} 2012, \aap, 548, A113

\bibitem[{{Downes} \& {Drury}(2014)}]{Downes-Drury-2014}
{Downes}, T.~P. \& {Drury}, L.~O. 2014, \mnras, 444, 365

\bibitem[{{Dyson}(1975)}]{Dyson-1975}
{Dyson}, J.~E. 1975, \apss, 35, 299

\bibitem[{{Edney}(1968)}]{Edney-1968}
{Edney}, B.~E. 1968, AIAA Journal, 6, 15

\bibitem[{Emanuel(2000)}]{Emanuel-2000}
Emanuel, G. 2000, in Handbook of Shock Waves, ed. T.~E. Gabi Ben-Dor,
  Ozer~Igra, Vol.~1 (Academic Press), 186--263

\bibitem[{{Fahr}(1990)}]{Fahr-1990}
{Fahr}, H.~J. 1990, \aap, 236, 86

\bibitem[{{Fahr} {et~al.}(2000){Fahr}, {Kausch}, \& {Scherer}}]{Fahr-etal-2000}
{Fahr}, H.~J., {Kausch}, T., \& {Scherer}, H. 2000, \aap, 357, 268

\bibitem[{{Fahr} {et~al.}(1995){Fahr}, {Scherer}, \&
  {Banaszkiewicz}}]{Fahr-etal-1995}
{Fahr}, H.~J., {Scherer}, K., \& {Banaszkiewicz}, M. 1995, \planss, 43, 301

\bibitem[{{Farris} \& {Russell}(1994)}]{Farris-Russell-1994}
{Farris}, M.~H. \& {Russell}, C.~T. 1994, \jgr, 99, 17681

\bibitem[{{Florinski} {et~al.}(2004){Florinski}, {Zank}, {Jokipii}, {Stone}, \&
  {Cummings}}]{Florinski-etal-2004}
{Florinski}, V., {Zank}, G.~P., {Jokipii}, J.~R., {Stone}, E.~C., \&
  {Cummings}, A.~C. 2004, \apj, 610, 1169

\bibitem[{{Goedbloed}(2008)}]{Goedbloed-2008}
{Goedbloed}, J.~P. 2008, Physics of Plasmas, 15, 062101

\bibitem[{{Goedbloed} {et~al.}(2010){Goedbloed}, {Keppens}, \&
  {Poedts}}]{Goedbloed-etal-2010}
{Goedbloed}, J.~P., {Keppens}, R., \& {Poedts}, S. 2010, {Advanced
  Magnetohydrodynamics} (Cambridge, UK: Cambridge University Press)

\bibitem[{{Goedbloed} \& {Poedts}(2004)}]{Goedbloed-Poedts-2004}
{Goedbloed}, J.~P.~H. \& {Poedts}, S. 2004, {Principles of
  Magnetohydrodynamics} (Cambridge University Press)

\bibitem[{{Gvaramadze} \& {Bomans}(2008)}]{Gvaramadze-Bomans-2008}
{Gvaramadze}, V.~V. \& {Bomans}, D.~J. 2008, \aap, 490, 1071

\bibitem[{{Gvaramadze} {et~al.}(2011){Gvaramadze}, {Kniazev}, {Kroupa}, \&
  {Oh}}]{Gvaramadze-etal-2011}
{Gvaramadze}, V.~V., {Kniazev}, A.~Y., {Kroupa}, P., \& {Oh}, S. 2011, \aap,
  535, A29

\bibitem[{{Heerikhuisen} {et~al.}(2006){Heerikhuisen}, {Florinski}, \&
  {Zank}}]{Heerikhuisen-etal-2006}
{Heerikhuisen}, J., {Florinski}, V., \& {Zank}, G.~P. 2006, \jgr, 111, 6110

\bibitem[{{Huthoff} \& {Kaper}(2002)}]{Huthoff-Kaper-2002}
{Huthoff}, F. \& {Kaper}, L. 2002, \aap, 383, 999

\bibitem[{{Izmodenov} {et~al.}(2003){Izmodenov}, {Malama}, {Gloeckler}, \&
  {Geiss}}]{Izmodenov-etal-2003a}
{Izmodenov}, V., {Malama}, Y.~G., {Gloeckler}, G., \& {Geiss}, J. 2003, \apjl,
  594, L59

\bibitem[{{Izmodenov} {et~al.}(2014){Izmodenov}, {Alexashov}, \&
  {Ruderman}}]{Izmodenov-etal-2014}
{Izmodenov}, V.~V., {Alexashov}, D.~B., \& {Ruderman}, M.~S. 2014, \apjl, 795,
  L7

\bibitem[{{Izmodenov} \& {Baranov}(2006)}]{Izmodenov-Baranov-2006}
{Izmodenov}, V.~V. \& {Baranov}, V.~B. 2006, ISSI Scientific Reports Series, 5,
  67

\bibitem[{{Jun} {et~al.}(1994){Jun}, {Clarke}, \& {Norman}}]{Jun-etal-1994}
{Jun}, B.-I., {Clarke}, D.~A., \& {Norman}, M.~L. 1994, \apj, 429, 748

\bibitem[{{Kissmann} {et~al.}(2008){Kissmann}, {Kleimann}, {Fichtner}, \&
  {Grauer}}]{Kissmann-etal-2008}
{Kissmann}, R., {Kleimann}, J., {Fichtner}, H., \& {Grauer}, R. 2008, \mnras,
  391, 1577

\bibitem[{{Kleimann} {et~al.}(2009){Kleimann}, {Kopp}, {Fichtner}, \&
  {Grauer}}]{Kleimann-etal-2009}
{Kleimann}, J., {Kopp}, A., {Fichtner}, H., \& {Grauer}, R. 2009, Annales
  Geophysicae, 27, 989

\bibitem[{{Kobulnicky} {et~al.}(2010){Kobulnicky}, {Gilbert}, \&
  {Kiminki}}]{Kobulnicky-etal-2010}
{Kobulnicky}, H.~A., {Gilbert}, I.~J., \& {Kiminki}, D.~C. 2010, \apj, 710, 549

\bibitem[{{Kosi{\'n}ski} \& {Hanasz}(2006)}]{Kosinski-Hanasz-2006}
{Kosi{\'n}ski}, R. \& {Hanasz}, M. 2006, \mnras, 368, 759

\bibitem[{{Linsky} \& {Wood}(2014)}]{Linsky-Wood-2014}
{Linsky}, J.~L. \& {Wood}, B.~E. 2014, ASTRA Proceedings, 1, 43

\bibitem[{{Liseau} {et~al.}(2015){Liseau}, {Larsson}, {Lunttila}, {Olberg},
  {Rydbeck}, {Bergman}, {Justtanont}, {Olofsson}, \& {de
  Vries}}]{Liseau-etal-2015}
{Liseau}, R., {Larsson}, B., {Lunttila}, T., {et~al.} 2015, \aap, 578, A131

\bibitem[{{Lowrie} {et~al.}(1999){Lowrie}, {Morel}, \&
  {Hittinger}}]{Lowrie-etal-1999}
{Lowrie}, R.~B., {Morel}, J.~E., \& {Hittinger}, J.~A. 1999, \apj, 521, 432

\bibitem[{{Mackey} {et~al.}(2015{\natexlab{a}}){Mackey}, {Gvaramadze},
  {Mohamed}, \& {Langer}}]{Mackey-etal-2015b}
{Mackey}, J., {Gvaramadze}, V.~V., {Mohamed}, S., \& {Langer}, N.
  2015{\natexlab{a}}, \aap, 573, A10

\bibitem[{{Mackey} {et~al.}(2015{\natexlab{b}}){Mackey}, {Gvaramadze},
  {Mohamed}, \& {Langer}}]{Mackey-etal-2015}
{Mackey}, J., {Gvaramadze}, V.~V., {Mohamed}, S., \& {Langer}, N.
  2015{\natexlab{b}}, \aap, 573, A10

\bibitem[{{Mackey} {et~al.}(2013){Mackey}, {Langer}, \&
  {Gvaramadze}}]{Mackey-etal-2013a}
{Mackey}, J., {Langer}, N., \& {Gvaramadze}, V.~V. 2013, \mnras, 436, 859

\bibitem[{{Mackey} {et~al.}(2014){Mackey}, {Langer}, {Mohamed}, {Gvaramadze},
  {Neilson}, \& {Meyer}}]{Mackey-etal-2014b}
{Mackey}, J., {Langer}, N., {Mohamed}, S., {et~al.} 2014, ASTRA Proceedings, 1,
  61

\bibitem[{{McComas} {et~al.}(2015){McComas}, {Bzowski}, {Fuselier}, {Frisch},
  {Galli}, {Izmodenov}, {Katushkina}, {Kubiak}, {Lee}, {Leonard}, {M{\"o}bius},
  {Park}, {Schwadron}, {Sok{\'o}{\l}}, {Swaczyna}, {Wood}, \&
  {Wurz}}]{McComas-etal-2015b}
{McComas}, D.~J., {Bzowski}, M., {Fuselier}, S.~A., {et~al.} 2015, \apjs, 220,
  22

\bibitem[{{Mellema} \& {Lundqvist}(2002)}]{Mellema-Lundqvist-2002}
{Mellema}, G. \& {Lundqvist}, P. 2002, \aap, 394, 901

\bibitem[{{Naca}(1953)}]{Naca-1135}
{Naca}. 1953, Equations, Tables and Charts for compressible flows, Tech. rep.,
  Ames Research staff, Ames Aeronautical Laboratory

\bibitem[{{Olivier}(2000)}]{Olivier-2000}
{Olivier}, H. 2000, Journal of Fluid Mechanics, 413, 345

\bibitem[{{Opher} {et~al.}(2012){Opher}, {Drake}, {Velli}, {Decker}, \&
  {Toth}}]{Opher-etal-2012}
{Opher}, M., {Drake}, J.~F., {Velli}, M., {Decker}, R.~B., \& {Toth}, G. 2012,
  \apj, 751, 80

\bibitem[{{Parker}(1961)}]{Parker-1961}
{Parker}, E.~N. 1961, \apj, 134, 20

\bibitem[{{Parkin} {et~al.}(2011){Parkin}, {Pittard}, {Corcoran}, \&
  {Hamaguchi}}]{Parkin-etal-2011}
{Parkin}, E.~R., {Pittard}, J.~M., {Corcoran}, M.~F., \& {Hamaguchi}, K. 2011,
  \apj, 726, 105

\bibitem[{{Pauls} {et~al.}(1995){Pauls}, {Zank}, \&
  {Williams}}]{Pauls-etal-1995}
{Pauls}, H.~L., {Zank}, G.~P., \& {Williams}, L.~L. 1995, \jgr, 100, 21595

\bibitem[{{Peri} {et~al.}(2012){Peri}, {Benaglia}, {Brookes}, {Stevens}, \&
  {Isequilla}}]{Peri-etal-2012}
{Peri}, C.~S., {Benaglia}, P., {Brookes}, D.~P., {Stevens}, I.~R., \&
  {Isequilla}, N.~L. 2012, \aap, 538, A108

\bibitem[{{Peri} {et~al.}(2015){Peri}, {Benaglia}, \&
  {Isequilla}}]{Peri-etal-2015}
{Peri}, C.~S., {Benaglia}, P., \& {Isequilla}, N.~L. 2015, \aap, 578, A45

\bibitem[{{Pogorelov} {et~al.}(2009){Pogorelov}, {Borovikov}, {Zank}, \&
  {Ogino}}]{Pogorelov-etal-2009}
{Pogorelov}, N.~V., {Borovikov}, S.~N., {Zank}, G.~P., \& {Ogino}, T. 2009,
  \apj, 696, 1478

\bibitem[{{Pogorelov} {et~al.}(2013){Pogorelov}, {Suess}, {Borovikov}, {Ebert},
  {McComas}, \& {Zank}}]{Pogorelov-etal-2013}
{Pogorelov}, N.~V., {Suess}, S.~T., {Borovikov}, S.~N., {et~al.} 2013, \apj,
  772, 2

\bibitem[{{Pogorelov} {et~al.}(2006){Pogorelov}, {Zank}, \&
  {Ogino}}]{Pogorelov-etal-2006}
{Pogorelov}, N.~V., {Zank}, G.~P., \& {Ogino}, T. 2006, \apj, 644, 1299

\bibitem[{{Povich} {et~al.}(2008){Povich}, {Benjamin}, {Whitney}, {Babler},
  {Indebetouw}, {Meade}, \& {Churchwell}}]{Povich-etal-2008}
{Povich}, M.~S., {Benjamin}, R.~A., {Whitney}, B.~A., {et~al.} 2008, \apj, 689,
  242

\bibitem[{{Raga} {et~al.}(2014){Raga}, {Cant{\'o}}, {Koenigsberger}, \&
  {Esquivel}}]{Raga-etal-2014}
{Raga}, A.~C., {Cant{\'o}}, J., {Koenigsberger}, G., \& {Esquivel}, A. 2014,
  \mnras, 443, 3284

\bibitem[{{Reitberger} {et~al.}(2014{\natexlab{a}}){Reitberger}, {Kissmann},
  {Reimer}, \& {Reimer}}]{Reitberger-etal-2014a}
{Reitberger}, K., {Kissmann}, R., {Reimer}, A., \& {Reimer}, O.
  2014{\natexlab{a}}, \apj, 789, 87

\bibitem[{{Reitberger} {et~al.}(2014{\natexlab{b}}){Reitberger}, {Kissmann},
  {Reimer}, {Reimer}, \& {Dubus}}]{Reitberger-etal-2014b}
{Reitberger}, K., {Kissmann}, R., {Reimer}, A., {Reimer}, O., \& {Dubus}, G.
  2014{\natexlab{b}}, \apj, 782, 96

\bibitem[{{Reynolds} {et~al.}(1999){Reynolds}, {Haffner}, \&
  {Tufte}}]{Reynolds-etal-1999}
{Reynolds}, R.~J., {Haffner}, L.~M., \& {Tufte}, S.~L. 1999, \apjl, 525, L21

\bibitem[{{Ritzerveld}(2005)}]{Ritzerveld-2005}
{Ritzerveld}, J. 2005, \aap, 439, L23

\bibitem[{{Rosner} {et~al.}(1978){Rosner}, {Tucker}, \&
  {Vaiana}}]{Rosner-etal-1978}
{Rosner}, R., {Tucker}, W.~H., \& {Vaiana}, G.~S. 1978, \apj, 220, 643

\bibitem[{{Salem} \& {Bryan}(2014)}]{Salem-Bryan-2014}
{Salem}, M. \& {Bryan}, G.~L. 2014, MNRAS, 437, 3312

\bibitem[{{Scherer} \& {Ferreira}(2005)}]{Scherer-Ferreira-2005a}
{Scherer}, K. \& {Ferreira}, S.~E.~S. 2005, ASTRA, 1, 17

\bibitem[{{Scherer} \& {Fichtner}(2014)}]{Scherer-Fichtner-2014}
{Scherer}, K. \& {Fichtner}, H. 2014, \apj

\bibitem[{{Scherer} {et~al.}(2014){Scherer}, {Fichtner}, {Fahr}, {Bzowski}, \&
  {Ferreira}}]{Scherer-etal-2014}
{Scherer}, K., {Fichtner}, H., {Fahr}, H.-J., {Bzowski}, M., \& {Ferreira}, S.
  2014, \aap

\bibitem[{{Scherer} {et~al.}(2015{\natexlab{a}}){Scherer}, {Fichtner}, {Fahr},
  \& {R\"oken}}]{Scherer-etal-2015c}
{Scherer}, K., {Fichtner}, H., {Fahr}, H.-J., \& {R\"oken}, C.
  2015{\natexlab{a}}, Submitted to \apj

\bibitem[{{Scherer} {et~al.}(2015{\natexlab{b}}){Scherer}, {van der Schyff},
  {Bomans}, {Ferreira}, {Fichtner}, {Kleimann}, {Strauss}, {Weis},
  {Wiengarten}, \& {Wodzinski}}]{Scherer-etal-2015a}
{Scherer}, K., {van der Schyff}, A., {Bomans}, D.~J., {et~al.}
  2015{\natexlab{b}}, \aap, 576, A97

\bibitem[{{Schneider}(1968)}]{Schneider-1968}
{Schneider}, W. 1968, Journal of Fluid Mechanics, 31, 397

\bibitem[{{Schulreich} \&
  {Breitschwerdt}(2011)}]{Schulreich-Breitschwerdt-2011}
{Schulreich}, M.~M. \& {Breitschwerdt}, D. 2011, \aap, 531, A13

\bibitem[{{Schure} {et~al.}(2009){Schure}, {Kosenko}, {Kaastra}, {Keppens}, \&
  {Vink}}]{Schure-etal-2009}
{Schure}, K.~M., {Kosenko}, D., {Kaastra}, J.~S., {Keppens}, R., \& {Vink}, J.
  2009, \aap, 508, 751

\bibitem[{{Schwarz} {et~al.}(1972){Schwarz}, {McCray}, \&
  {Stein}}]{Schwarz-etal-1972}
{Schwarz}, J., {McCray}, R., \& {Stein}, R.~F. 1972, \apj, 175, 673

\bibitem[{{Sexton} {et~al.}(2015){Sexton}, {Povich}, {Smith}, {Babler},
  {Meade}, \& {Rudolph}}]{Sexton-etal-2015}
{Sexton}, R.~O., {Povich}, M.~S., {Smith}, N., {et~al.} 2015, \mnras, 446, 1047

\bibitem[{{Siewert} {et~al.}(2004){Siewert}, {Pohl}, \&
  {Schlickeiser}}]{Siewert-etal-2004}
{Siewert}, M., {Pohl}, M., \& {Schlickeiser}, R. 2004, \aap, 425, 405

\bibitem[{{Sok{\'o}{\l}} {et~al.}(2015){Sok{\'o}{\l}}, {Bzowski}, {Kubiak},
  {Swaczyna}, {Galli}, {Wurz}, {M{\"o}bius}, {Kucharek}, {Fuselier}, \&
  {McComas}}]{Sokol-etal-2015}
{Sok{\'o}{\l}}, J.~M., {Bzowski}, M., {Kubiak}, M.~A., {et~al.} 2015, \apjs,
  220, 29

\bibitem[{{Steinolfson}(1994)}]{Steinolfson-1994}
{Steinolfson}, R.~S. 1994, \jgr, 99, 13307

\bibitem[{{Suess} \& {Nerney}(1990)}]{Suess-Nerney-1990}
{Suess}, S.~T. \& {Nerney}, S. 1990, \jgr, 95, 6403

\bibitem[{{Sutherland} \& {Dopita}(1993)}]{Sutherland-Dopita-1993}
{Sutherland}, R.~S. \& {Dopita}, M.~A. 1993, \apjs, 88, 253

\bibitem[{{Townsend}(2009)}]{Townsend-2009}
{Townsend}, R.~H.~D. 2009, \apjs, 181, 391

\bibitem[{{Usmanov} \& {Goldstein}(2006)}]{Usmanov-Goldstein-2006}
{Usmanov}, A.~V. \& {Goldstein}, M.~L. 2006, Journal of Geophysical Research
  (Space Physics), 111, 7101

\bibitem[{{Usmanov} {et~al.}(2014){Usmanov}, {Goldstein}, \&
  {Matthaeus}}]{Usmanov-etal-2014}
{Usmanov}, A.~V., {Goldstein}, M.~L., \& {Matthaeus}, W.~H. 2014, \apj, 788, 43

\bibitem[{Van~Dyke(1958)}]{van-Dyke-1958}
Van~Dyke, M.~D. 1958, Journal of Fluid Mechanics, 3, 515

\bibitem[{{van Leeuwen}(2007)}]{van-Leeuuwen-2007}
{van Leeuwen}, F. 2007, \aap, 474, 653

\bibitem[{{van Marle} {et~al.}(2014){van Marle}, {Decin}, \&
  {Meliani}}]{van-Marle-etal-2014}
{van Marle}, A.~J., {Decin}, L., \& {Meliani}, Z. 2014, \aap, 561, A152

\bibitem[{{van Marle} {et~al.}(2011){van Marle}, {Meliani}, {Keppens}, \&
  {Decin}}]{van-Marle-etal-2011}
{van Marle}, A.~J., {Meliani}, Z., {Keppens}, R., \& {Decin}, L. 2011, \apjl,
  734, L26

\bibitem[{{Wiengarten} {et~al.}(2015){Wiengarten}, {Fichtner}, {Kleimann}, \&
  {Kissmann}}]{Wiengarten-etal-2015}
{Wiengarten}, T., {Fichtner}, H., {Kleimann}, J., \& {Kissmann}, R. 2015, \apj,
  805, 155

\bibitem[{{Wood} {et~al.}(2007){Wood}, {Izmodenov}, {Linsky}, \&
  {Alexashov}}]{Wood-etal-2007}
{Wood}, B.~E., {Izmodenov}, V.~V., {Linsky}, J.~L., \& {Alexashov}, D. 2007,
  \apj, 659, 1784

\bibitem[{{Zank}(1999)}]{Zank-1999}
{Zank}, G.~P. 1999, \ssr, 89, 413

\bibitem[{{Zank} \& {Frisch}(1999)}]{Zank-Frisch-1999}
{Zank}, G.~P. \& {Frisch}, P.~C. 1999, \apj, 518, 965

\bibitem[{{Zank} {et~al.}(2013){Zank}, {Heerikhuisen}, {Wood}, {Pogorelov},
  {Zirnstein}, \& {McComas}}]{Zank-etal-2013}
{Zank}, G.~P., {Heerikhuisen}, J., {Wood}, B.~E., {et~al.} 2013, \apj, 763, 20

\end{thebibliography}

\end{document}